\documentclass[a4paper,twocolumn,11pt,accepted=2026-10-20]{quantumarticle}

\pdfoutput=1
\usepackage[utf8]{inputenc}
\usepackage[english]{babel}
\usepackage[T1]{fontenc}
\usepackage{amsmath}
\usepackage{hyperref}
\usepackage[sort&compress,numbers]{natbib}
\usepackage{tikz}
\usepackage{lipsum}
\usepackage{pythonhighlight}
\usepackage{braket}

\hypersetup{colorlinks=true, citecolor=blue, urlcolor=teal, linkcolor=blue}

\usepackage{xcolor}

\newenvironment{DIFnomarkup}{\ifdefined\DIFadd\color{blue}\fi}{}

\providecommand{\onecolumngrid}{}
\providecommand{\twocolumngrid}{}

\makeatletter
\newcommand{\blfootnote}[1]{%
  \begingroup
    \renewcommand{\thefootnote}{}%
    \renewcommand{\@makefnmark}{}%
    \footnote{#1}%
    \addtocounter{footnote}{-1}%
  \endgroup
}
\makeatother

\usepackage{tabularray}

\begin{document}

\title{QwaveMPS: An efficient open-source Python package for simulating non-Markovian waveguide-QED using matrix product states
}

\author{\texorpdfstring{Sofia Arranz Regidor\textsuperscript{*}}{Sofia Arranz Regidor}}
\orcid{0000-0002-7295-604X}
\email{18sar4@queensu.ca}

\author{\texorpdfstring{Matthew Kozma\textsuperscript{*}}{Matthew Kozma}}
\orcid{0009-0005-2296-2790}
\email{m.kozma@queensu.ca}

\author{Stephen Hughes}
\orcid{0000-0002-5486-2015}
\email{shughes@queensu.ca}
\affiliation{Department of Physics, Engineering Physics and Astronomy, Queen's University, Kingston ON K7L 3N6, Canada}

\maketitle
\blfootnote{$^{*}$These two authors contributed equally to this work.}

\begin{abstract}
  QwaveMPS is an open-source Python library for simulating one-dimensional quantum many-body waveguide systems using matrix product states (MPS). It provides a user-friendly interface for constructing, evolving, and analyzing quantum states and operators, facilitating studies in quantum physics and quantum information with waveguide-QED systems. This approach enables efficient, scalable simulations by focusing computational resources on the most relevant parts of the quantum system. Thus, one can study
  a wide range of complex dynamical interactions, including time-delayed feedback effects in the non-Markovian regime and deeply nonlinear systems, at a highly reduced computational cost compared to full Hilbert space approaches, making it both practical and convenient to model a variety of open waveguide-QED systems (in Markovian and non-Markovian regimes), treating quantized atoms and quantized photons on an equal footing. The
  QwaveMPS package automates the waveguide-QED-specific tensor-network operations, including construction of the time-bin evolution operator from physical parameters, the swap network required for delayed interactions, orthogonality-center and truncation management, the matrix-product factorization of $N$-photon Fock-state inputs, and the evaluation of single-time and two-time field observables (such as the second-order quantum correlation function) at chosen points along the waveguide.
\end{abstract}


\begin{DIFnomarkup}
\href{https://github.com/QwaveMPS/QwaveMPS}{github.com/QwaveMPS/QwaveMPS}
\end{DIFnomarkup}

\section{Introduction}
\begin{figure*}
    \centering
    \includegraphics[width=\textwidth]{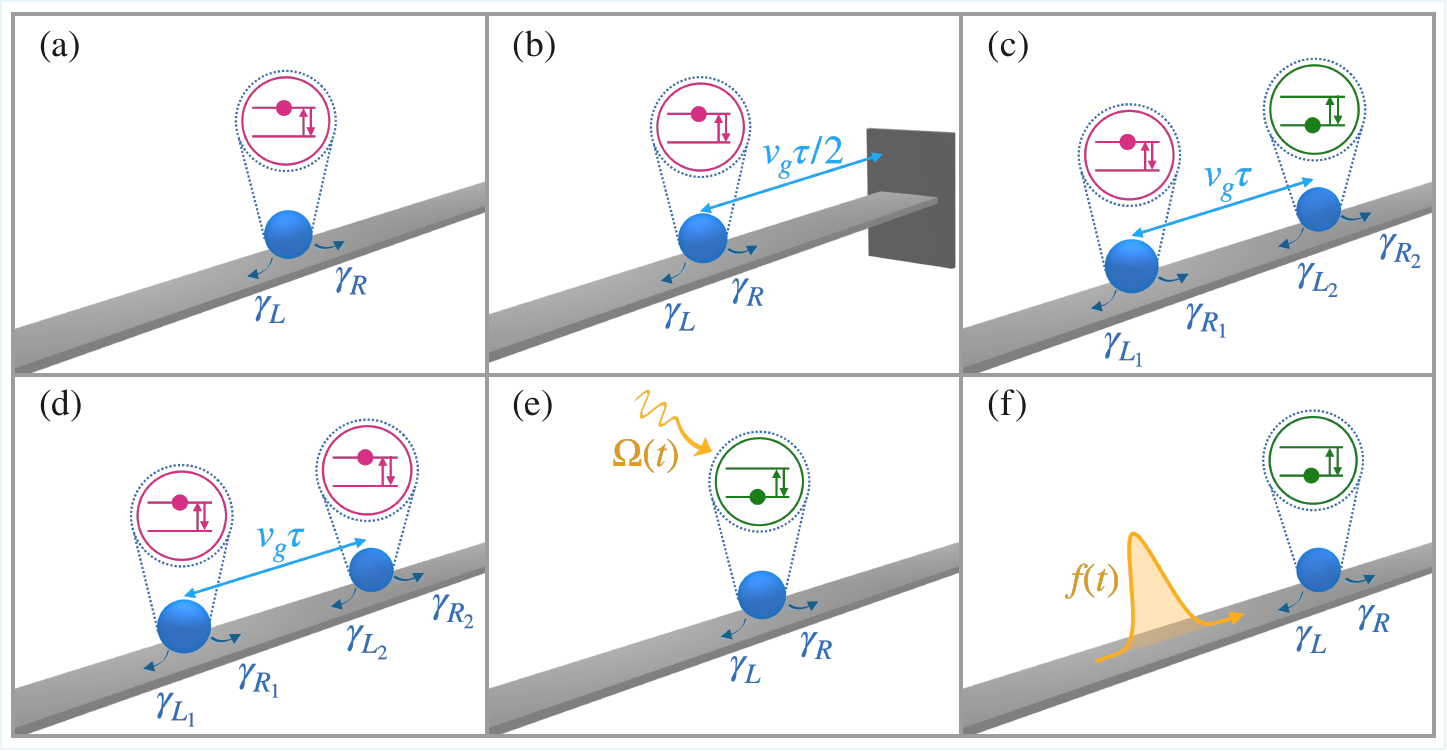}
    \caption{Schematic of some example cases calculated using QwaveMPS. These can represent various waveguide-QED systems, including semiconductors and superconducting circuits. (a) Decay of a TLS in a waveguide, where $\gamma_L$ and $\gamma_R$ are the left/right coupling rates. (b) Decay of a TLS in a semi-infinite waveguide with a side mirror at a distance $d= v_g \tau /2$, with $v_g$ the group velocity and $\tau$ the roundtrip delay time. (c) Decay of two TLSs, where one is initially excited, and the other one is in the ground state (linear regime). (d) Decay of two TLSs initially excited (nonlinear regime). (e) A TLS in a waveguide perpendicularly driven by a time-dependent classical field $\Omega(t)$. (f) Single TLS driven by a quantum pulse (such as an $N$-photon Fock state) with an envelope shape $f(t)$. 
    }  
    \label{schematic}
\end{figure*}

Waveguide quantum electrodynamics (waveguide-QED) is a platform for studying
quantum light-matter interactions in quasi-one-dimensional geometries, where
quantum emitters---atoms, quantum dots, or superconducting qubits---couple
strongly to a continuum of quantized field modes~\cite{PhysRevA.76.062709,PhysRevLett.98.153003,Witthaut_2010,PhysRevLett.106.053601,PhysRevLett.113.263604,
PhysRevLett.116.093601,Calaj2016,Hughes2004,PhysRevLett.120.140404,PhysRevA.87.013820,PhysRevLett.118.213601,PhysRevA.82.063816,PhysRevA.83.063828,PhysRevA.102.023702,PhysRevA.100.023834,Dinc2019,PhysRevLett.122.073601,GonzalezBallestero2013,RevModPhys.89.021001,Kannan2020,doi:10.1126/sciadv.aaw0297,PhysRevResearch.6.L032017,Lodahl2017,RevModPhys.95.015002}. This opens up a wide
range of applications in quantum optics and quantum technologies~\cite{BROWNE20172,Wang2025},
where photons and atoms interact strongly.

As an example application, having quantum emitters such as two-level systems (TLSs) or superconducting transmons in a waveguide gives rise to interesting effects such as time-delayed feedback phenomena (yielding non-Markovian effects) and long-lived vacuum quantum oscillations through controlled feedback, with the ability to continuously re-absorb and re-emit photons by the emitters. Additionally, the local symmetry of the problem can be broken by using chiral systems, where the emitters couple to photons only in one waveguide direction~\cite{gardiner_zoller_2010,PhysRevResearch.3.023030,PhysRevA.106.023708}.

Modeling such systems in an exact quantum mechanical way, however, poses a significant challenge. Their rapidly increasing Hilbert space, combined with strong nonlinearities and long-time delays, presents a substantial obstacle for accurately solving these types of problems. 
Thus, most approaches make use of the common {\it Markov approximation}, in which the delayed feedback effects are disregarded, and all coupling rates are instantaneous. Examples of such approaches include the master equation approach and input-output theories; although these highly simplify the problem and are widely used, there are 
various regimes where they fail (and often drastically), and non-Markovian effects must be taken into consideration.

Other theory approaches can extend beyond the Markov approximation, such as scattering theories and quantum stochastic methods~\cite{PhysRevA.82.063821,Rephaeli2012FewPhotonSC,PhysRevA.101.023807,Nysteen2015,Chen_2011,chen2026resonancefluorescenceartificialatom}; however, these methods are typically restricted to weak excitations or low photon numbers, which is not suitable for modeling strongly nonlinear time-delayed regimes. Recent progress in superconducting quantum circuits~\cite{You2011,blais_circuit_2021,PhysRevX.14.031055,Marcaud2025,7rnm-rxhh,PhysRevResearch.5.033155} and quantum nanophotonics~\cite{LeJeannic2022,PhysRevLett.126.023603,Patel2026single} has intensified the demand for numerically exact and scalable simulation techniques capable of capturing the full quantum dynamics of complex quantum circuits.
In superconducting systems, dipoles coupled to different parts of a complex waveguide can also realize so-called giant atom effects
~\cite{PhysRevA.106.063717,PhysRevLett.120.140404,PhysRevResearch.2.043184,PhysRevResearch.2.043070,Kannan2020,Wang_2022,Liu:22,PhysRevResearch.6.013279}.

In our waveguide-QED software package, we exploit a matrix product state (MPS) method to address these common model limitations. 
For quantum circuits with delay lines and quantum feedback, the approach we use was pioneered by Pichler and Zoller~\cite{PhysRevLett.116.093601,PhysRevA.93.062104}.
It is important to note that the
MPS formalism is a numerically exact approach, based on tensor network (TN) theories, that allows one to discretize the electromagnetic field into time bins, in order to limit the growth of the Hilbert space and solve the problem in a tractable manner~\cite{hmbj-lg7p,PhysRevX.14.031043,Capurso2026}. It was originally developed for computing strongly correlated systems, such as finite-size coupled spin chains, and it has been successfully used by leading companies to benchmark their most advanced superconducting quantum processors~\cite{mcculloch_density-matrix_2007,Jaderberg:2025iei,PhysRevA.109.062437}.

Methods based on MPS have been successfully employed to model various quantum circuits
and waveguides, including non-Markovian environments, time-delayed feedback, and strongly correlated photonic states~\cite{Manzoni2017}. However, it remains a relatively niche tool within the broader quantum optics community, in contrast to other approaches such as master equation techniques, which have been widely used in software frameworks such as the excellent {\em QuTiP}~\cite{johansson2012qutip,johansson_qutip_2013} package, or waveguide-QED approaches restricted to low photon numbers, such as the {\em WaveguideQED.jl} package~\cite{BundgaardNielsen2025waveguideqedjl}, which is limited to solve problems with a maximum of two quanta, and semi-analytical solutions~\cite{Dinc2019exactmarkoviannon}. This is largely due to its technical complexity and the lack of specific documentation and accessible simulation tools in the quantum optics framework for quantum circuits and waveguide-QED.

In this work, we introduce QwaveMPS, our open-source Python library designed to simulate waveguide-QED systems using MPS. {QwaveMPS} combines a user-friendly interface with TN algorithms, enabling efficient and scalable simulations of open one-dimensional quantum systems with delayed feedback, strong nonlinearities, and multiple propagating quanta. By making the use of MPS methods more accessible in quantum optics and quantum circuits, QwaveMPS aims to bridge the gap between powerful theoretical techniques and practical and reproducible studies. Importantly, users can use and benefit from {\it QwaveMPS} without requiring a detailed knowledge of TNs. This allows users to avoid having to build from scratch a time-bin waveguide-QED simulation model, time evolution, and observable evaluation on top of a general-purpose TN library.


We begin by presenting the basic theoretical background of MPS in Sec.~\ref{sec:theory}, where the principles of MPS applied to waveguide-QED systems are explained; we introduce the fundamentals of MPS, show a useful diagrammatic representation of states and operators, and discuss the time evolution of the system. Then, in Sec.~\ref{sec:package}, we introduce QwaveMPS, our open-source Python package, with an overview of the package framework and usage. QwaveMPS performs steps from time-bin evolution to field observables internally, and Sec.~\ref{subsec:automation} sets out exactly which ones. Example cases showing its potential both in the Markovian and non-Markovian regimes are shown in Sec.~\ref{sec:examples}. These examples are pictured schematically in Fig.~\ref{schematic}. We start with the simplest case of vacuum dynamics in the linear regime in Sec.~\ref{subsec:linear_regime} 
[Figs.~\ref{schematic}(a)--\ref{schematic}(c)], and then introduce nonlinear dynamics and quantum correlations in Sec.~\ref{subsec:vac_nonlinear} [Fig.~\ref{schematic}(d)]. Section~\ref{subsec:classical_pump} introduces the use of external classical drives with continuous-wave pumps and pulsed light [Fig.~\ref{schematic}(e)]. Afterwards, dynamics with full quantum Fock state pulses are solved in Sec.~\ref{subsec:fock} [Fig.~\ref{schematic}(f)]. 

Although we already address here a fairly broad range of important problems in waveguide-QED quantum optics, MPS is not limited to these cases, and in Sec.~\ref{subsec:future}, we discuss some features that could be implemented in future versions of the QwaveMPS package. Finally, in Sec.~\ref{sec:conclusion} we give some concluding remarks.

\section{Theoretical background}
\label{sec:theory}

\subsection{Waveguide-QED}

Waveguide-QED deals with quasi-one-dimensional quantum systems where atoms or quantum emitters couple to a continuum of quantized field modes. These systems show rich light-matter phenomena that are unique to the waveguide geometry, and can include significant non-Markovian or delayed feedback effects, which can lead to the emission and complete reabsorption of photons by the quantum emitters, or the ability to couple to chiral emitters, breaking the local symmetry of the problem by emitting photons only in one waveguide direction for a better flow of the quantum information. Resonant single photons can also be scattered and reflected elastically, or completely invert a TLS~\cite{PhysRevA.82.033804}. 

The following theory is in units of $\hbar=1$. The general form of the total Hamiltonian is
\begin{equation}
    H = H_{\rm sys} + H_{\rm W} +  H_{\rm I},
\end{equation}
where $H_{\rm sys}$ corresponds to the emitters' 
Hamiltonian: 
\begin{equation}
    H_{\rm sys} = \sum_n \left[ \omega_n \sigma_n^+ \sigma_n^- {-} \frac{1}{2} \left( \Omega_n(t) \sigma_n^- e^{i\omega_L t} {+} \rm{H.c.} \right) \right],
\end{equation}
 which can also include a classical pump $\Omega_n(t)$ centered at a frequency $\omega_L$. 
 In addition, 
$H_{\rm W}$ is the waveguide Hamiltonian, and $H_{\rm I}$ represents the interaction between the emitters (one or more) and the waveguide modes of interest. The waveguide Hamiltonian can be expressed as,
\begin{equation}
    H_{\rm W} = \sum_{\alpha= L,R} \int_{\mathcal{B}} d\omega  \omega b_\alpha^\dagger (\omega)b_\alpha(\omega), 
\end{equation}
where $b_\alpha^\dagger (\omega)$ and $b_\alpha(\omega)$ are the bosonic creation and annihilation operators for the right- and left-moving photons, and $\mathcal{B}$ is the relevant bandwidth of interest around the resonance frequency $\omega_0$, in the rotating-wave approximation.

The interaction Hamiltonian between the field and $n$ emitters is 
\begin{equation}
\begin{split}
    H_{\rm I} &= \sum_{n}  \sum_{\alpha= L,R} \frac{i}{\sqrt{2\pi}}\int_{\mathcal{B}} d \omega \ \big[ \kappa^*_{n,\alpha} (\omega)  \sigma^-_n b_\alpha^\dagger(\omega) \\
   &- \kappa_{n,\alpha} (\omega) b_\alpha(\omega) \sigma^+_n \big],
\end{split}    
\label{eq:Hint}
\end{equation}
where $\kappa_{n,\alpha}$ is the coupling rate, which is assumed to be frequency independent (in the bandwidth of interest). Thus, we can replace
$\kappa_{n,\alpha}(\omega)$ by $\kappa_{n,\alpha}(\omega_0)=\sqrt{{\gamma_{n,\alpha}}/{2\pi}}$, with $\gamma_{n,\alpha}$ being the decay rate of the $n$th emitter.

Then, we can transform the Hamiltonian to the time domain by defining a time-dependent noise operator in terms of the field operators:
\begin{equation}
    b_\alpha(t) = \int \frac{1}{\sqrt{2\pi}} d \omega \ b_\alpha(\omega) e^{-i (\omega - \omega_0) t},
\end{equation}
with the white noise commutation relation $[b_\alpha(t),b_{\alpha'}^\dagger(t')] = \delta(t-t')\delta_{\alpha,\alpha'}$. We can then rewrite the interaction Hamiltonian as
\begin{equation}
\begin{split}
    H_{\rm I} &= \sum_{n}  \sum_{\alpha= L,R} i \big[ \kappa^*_{n,\alpha}  \sigma^-_n b_\alpha^\dagger(t-x_n/v) \\
   &- \kappa_{n,\alpha} b_\alpha(t-x_n/v) \sigma^+_n \big],
\end{split}    
\label{eq:Hint_t}
\end{equation}
where $x_n$ corresponds to the position of the atom, and $v$ is the velocity of propagation. Here, we are working in the interaction picture, where $\sigma_n^- (t) = \sigma_n^- \ e^{-i\omega_0 t}$, and
$b_\alpha(\omega,t) = b_\alpha(\omega)  e^{-i \omega t}$.
 
To solve the general class of complex
waveguide-QED problems mentioned earlier, many traditional quantum optics theories make use of approximations, such as the Markovian approximation, which results in potentially valuable information being lost (including quantum coherence, causality effects and entanglement). 
In contrast, MPS allows one to solve waveguide-QED systems
 without invoking the usual approximations~\cite{PhysRevLett.116.093601,PhysRevResearch.3.023030,Pichler11362}. Moreover, we stress that the general MPS methods are computationally efficient as well as
{\it numerically exact}.  

For the waveguide-based MPS approach, we need to discretize the waveguide in time into so-called {\it time bins}. One can then write the discretized Hamiltonian in time, using the boson noise operators~\cite{PhysRevLett.116.093601},
\begin{equation}
    \Delta B_{\alpha}^{(\dagger)}(t_k)  = \int_{t_k}^{t_{k+1}} dt' b_{\alpha}^{(\dagger)}(t'),
    \label{eq:noise_op}
\end{equation}
where these operators create/annihilate a photon in a time bin and have the commutation relations $\left[ \Delta B_{\alpha}(t_k), \Delta B_{\alpha'}^\dagger(t_{k'}) \right] = \Delta t \delta_{k,k'} \delta_{\alpha,\alpha'}$, with $\alpha=R,L$ for the right- and left-moving photons, respectively, and $\Delta t$ being the time bin resolution
(discretized time step). 

With this definition, we create a new basis,
\begin{equation}
    \ket{i^\alpha_k} = \frac{(\Delta B_\alpha^\dagger (t_k))^{i^\alpha_k}}{\sqrt{i^\alpha_k ! (\Delta t)^{i^\alpha_k}}} \ket{\rm vac},
\end{equation}
and write the time evolution operator for each time step: 
\begin{equation}
    U(t_{k+1},t_k) =  \exp{ \left( -i \int_{t_k}^{t_{k+1}} dt' H(t')\right)}.
    \label{eq:t_evol}
\end{equation}
Then, the initial state is written in terms of a matrix product state, and the time evolution operator is transformed to a matrix product operator (MPO) 
so we can evolve the entire system one time step at a time. 

\subsection{Matrix Product States}
\label{subsec:mps}
\subsubsection{Singular value decomposition}

\begin{figure}[ht]
  \centering
  \includegraphics[width=0.7\columnwidth]{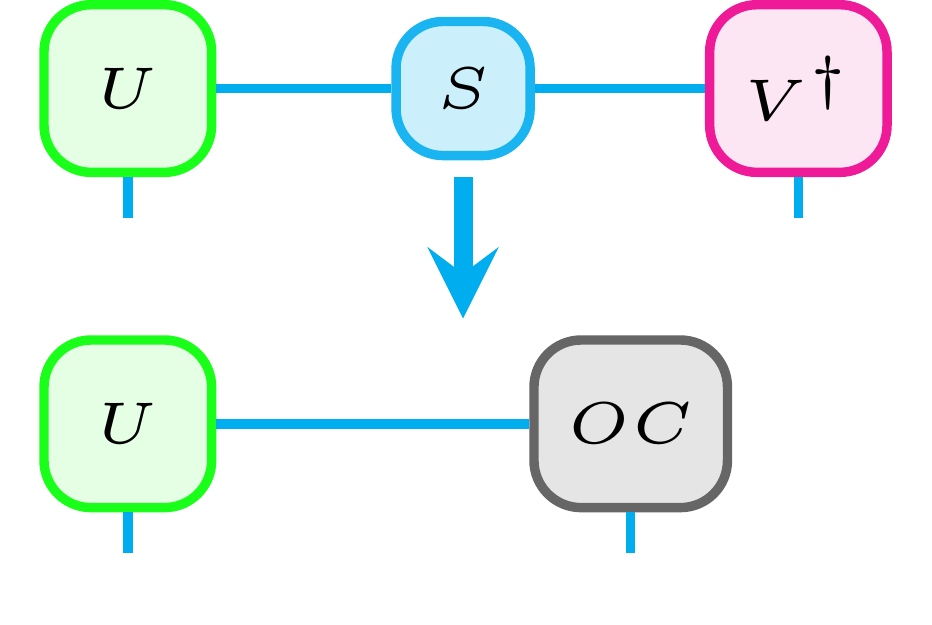}
  \caption{Diagrammatic representation of an SVD, where $U$ represents a left-normalized tensor (green bins), $V^\dag$ is a right-normalized tensor (magenta bin) and $S$ represents the diagonal matrix with the Schmidt coefficients (blue bin). The lower diagram shows the subsequent contraction of the Schmidt coefficients with $V^\dag$, where $OC$ represents the orthogonality center (grey bin).}
  \label{fig:oc}
\end{figure}

Matrix product states is an approach based on one-dimensional tensor network theories~\cite{PhysRevResearch.3.023030,orus_practical_2014}. The MPS algorithm relies on the Schmidt decomposition or singular value decomposition (SVD) of a quantum system, which considers the bipartition state of the system as a tensor product. The SVD of a tensor $M$ is
\begin{equation}
    M = U S V^\dagger,
\end{equation}
where $S$ is a diagonal matrix containing the Schmidt coefficients in descending order, $U$ is a left-normalized tensor, and $V^\dag$ is a right-normalized tensor. Afterwards, one of the side tensors can be contracted with the tensor containing the Schmidt coefficients. This is called the orthogonality center (OC) and carries the system's information. Thus, we end up with two new tensors written as a tensor product. To better understand the process, this is represented diagrammatically in Fig.~\ref{fig:oc}.

Here, the vertical lines correspond to the physical dimensions of the system, while the horizontal ones represent the bond or virtual extra dimensions generated when performing the SVDs. The new bond dimensions will carry the entanglement of the system, and their size will depend on the level of entanglement generated in each case. This entanglement will increase with the number of excitations/photons considered in the system, as well as with the delayed feedback. 

\subsubsection{Matrix product states}

By iterating this process, we can decompose the Hilbert space into a tensor product of smaller subspaces until we get the following general MPS expression for a waveguide-QED system,
\begin{equation}
\begin{split}
       &\ket{\psi}= \\
       &\sum_{i_s i_1...i_N} A_{a_1}^{i_s}A_{a_1,a_2}^{i_1} ... A_{a_{N-1},a_{N}}^{i_{N-1}}A_{a_{N}}^{i_{N}}
   \ket{i_s, i_1,...,i_N}, 
\end{split}
\end{equation}
where the first term represents the system (or quantum emitter) part, and the remaining $N$ terms represent the waveguide discretized in time. 

\begin{figure}[ht]
  \centering
  \includegraphics[width=\columnwidth]{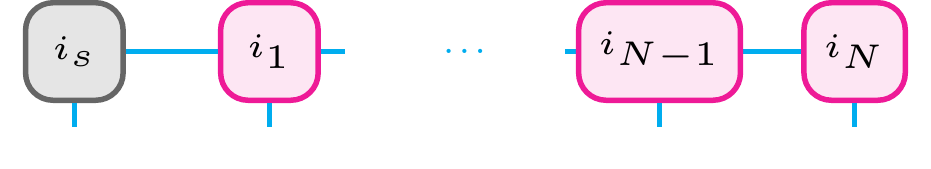}
  \caption{Diagrammatic representation of a state written as a matrix product state. 
  The grey bin represents the emitter subspace and contains the OC, and the magenta bins are right-normalized tensors containing the field discretized in time, with each bin corresponding to a time step. In this diagram, $i_s i_1...i_N$ represent the physical indices of the MPS.}
  \label{fig:init_state}
\end{figure}

Here, each tensor can be represented as a `bin' which corresponds to the boxes in the diagrammatic representation in Fig.~\ref{fig:init_state}. By changing the dimension of the time bins, one can include probabilities for one or more photons per time step. This gives the possibility of having denser probabilities of photons, as is necessary for simulating Fock states and calculating higher-order field correlations.

\subsubsection{Matrix product operators}

An operator can be seen as a map that takes one physical index $i$ to another $j$ with some coefficients $O^{ij}$. Thus, MPOs have two physical indices per site. The main advantage is that the whole state does not need to be computed when an operator is applied. Only the sites involved in the operation are computed, highly reducing the computational cost of the operation.

\begin{figure}[ht]
  \centering
  \includegraphics[width=0.2\columnwidth]{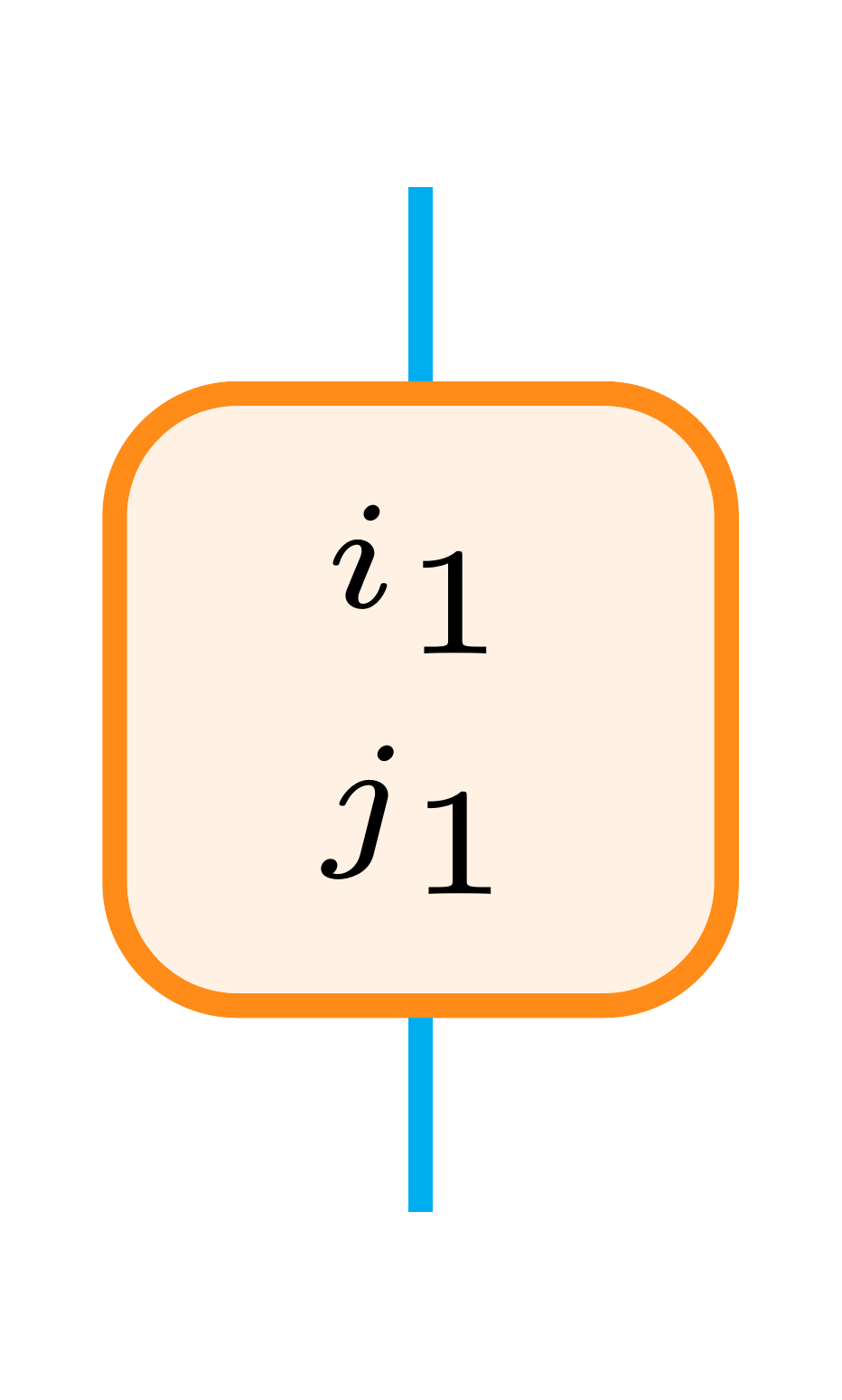}
  \caption{Diagrammatic representation of a single site operator written as an MPO. Here, $i_1$ and $j_1$ represent the physical indices of the MPO. }
  \label{fig:mpo}
\end{figure}

For example, an MPO operating on a single site can be represented as in Fig.~\ref{fig:mpo}, where $i_1$ and $j_1$ are the labels for the physical indices of the corresponding bra and ket. This can be directly applied on the corresponding MPS bin.

\subsubsection{Time evolution}

\begin{figure}[ht]
  \centering
  \includegraphics[width=0.4\columnwidth]{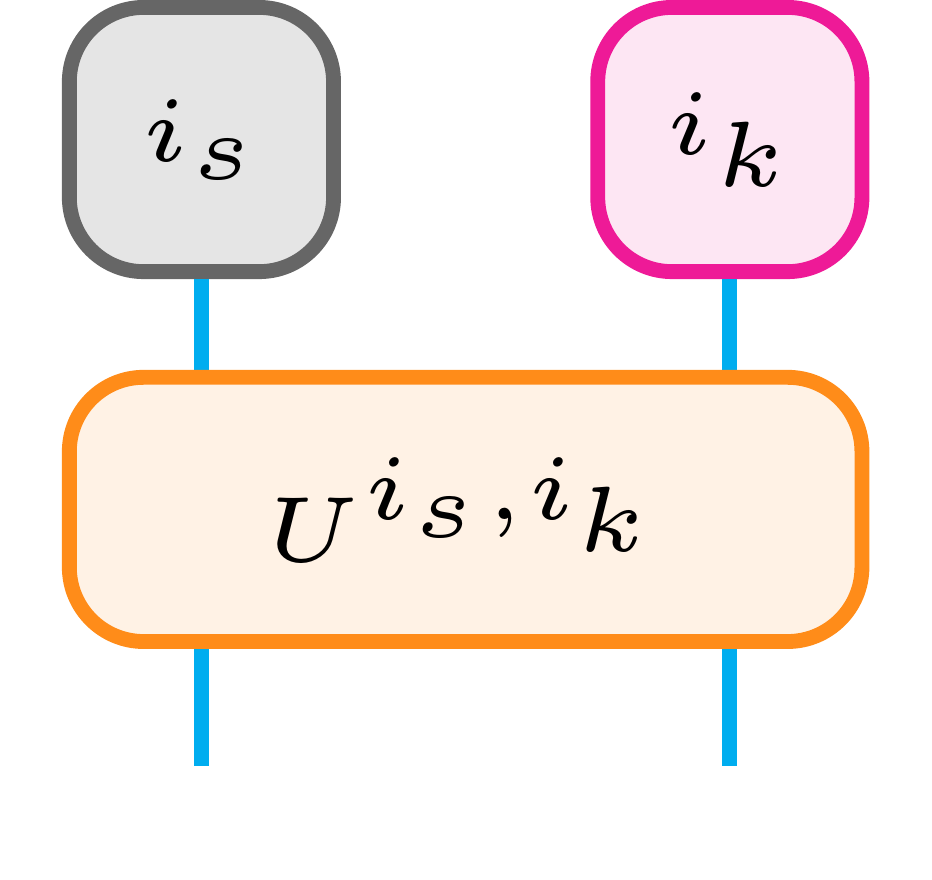}
  \caption{Diagrammatic representation of a Markovian time evolution operator $U$ operating at a time step $t_k$ on the system bin (grey bin) and the corresponding present bin (magenta bin). Here, $i_s$ is the system physical index and $i_k$ is the time bin one. }
  \label{fig:mark_u}
\end{figure}

The evolution of the system is performed by applying the time evolution operator, $U$, on the relevant parts of the MPS at each time step. In the Markovian regime or when we do not have feedback effects, this is usually on the system bin and the present time bin. For example, at a time $t_k$, $U$ will operate on the system bin with physical index $i_s$ and the time bin with physical index $i_k$ as in Fig.~\ref{fig:mark_u}. In this case, the evolution operator is a 2-site MPO. 

\begin{figure}[t]
  \centering
  \includegraphics[width=\columnwidth]{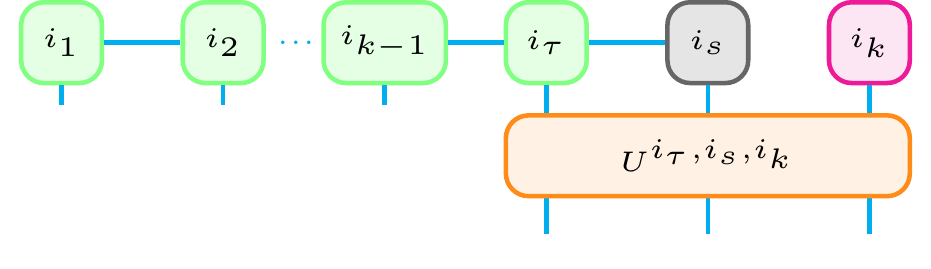}
  \caption{Diagrammatic representation of a non-Markovian time evolution operator $U$ operating at a time step $t_k$ on the system bin (grey bin), the corresponding present bin (magenta bin) and the feedback bin at $\tau$ (green bin). Here, $i_s$ is the system physical index and $i_k$ is the time bin index, and $i_\tau$ is the feedback one.}
  \label{fig:nomark_u}
\end{figure}

However, in cases with feedback effects in the non-Markovian regime (e.g., if including backscattering from emitters or a mirror), we will have to consider also the feedback bins (time delays and retardation). For example, if we have a feedback bin with physical index $i_\tau$, the time evolution operator is now a 3-site MPO as in Fig.~\ref{fig:nomark_u}.

Here, we have first brought the feedback time bin next to the system bin $i_s$ and the present time bin $i_k$ by applying a swap operator, in order to then apply $U$ only in three sites. For a time delay of $\tau$, the chain of feedback bins is of length ${\rm round}\left(\frac{\tau}{\Delta t}\right)$, and requires a proportional number of swaps to place the bins in the appropriate locations in the MPS chain.

In all cases, after applying the time evolution operator, the state is again decomposed, performing SVD to recover the MPS shape of each bin. By iterating this process, the evolution of the system is calculated.

\section{QwaveMPS package overview}
\label{sec:package}

\begin{figure*}[t]
  \centering
  \includegraphics[width=\textwidth]{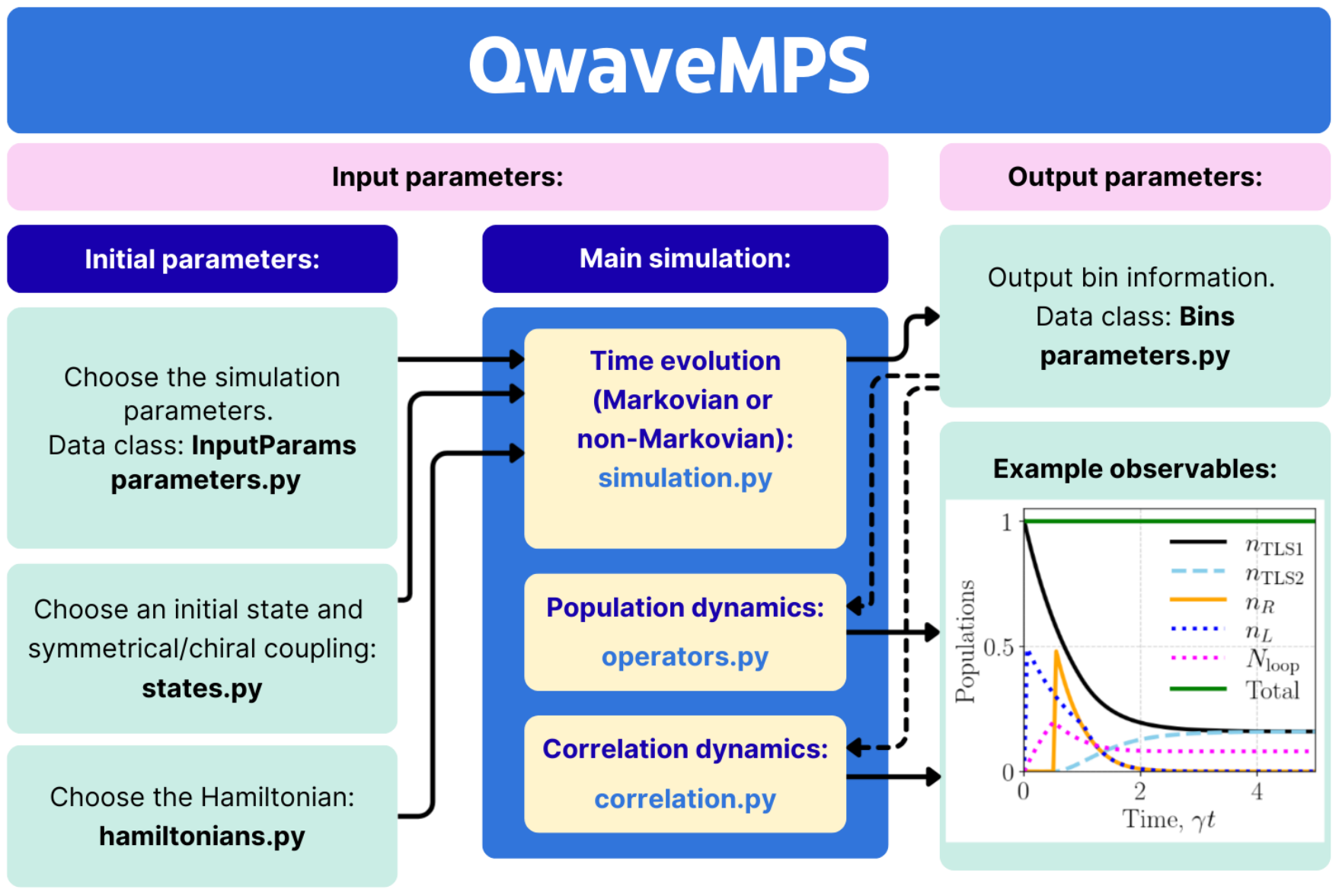}
  \caption{Diagram showing an example of the package framework. The input parameters include the initial parameters needed to set up the system and run the main simulation, where the time evolution of the total system state is calculated (stored in the \textit{Bins} data class). Observables such as TLS populations and field fluxes can be calculated by taking the expectation values of the relevant operators on the appropriate emitter/field bins. An example output plot with some observables in the non-Markovian regime is included in the bottom right.}
  \label{fig:diagram}
\end{figure*}

In this section, we introduce the QwaveMPS package, its implementation, and usage. QwaveMPS is a Python package that allows one to solve waveguide-QED problems using the MPS techniques introduced in the previous section. The source code is publicly available on GitHub~\cite{githubQwaveMPS}. 

Figure~\ref{fig:diagram} contains a diagram with the basic components of the package. There are six source scripts in the package. Three of them contain information for the initial parameters: these are the possible initial states (\texttt{states.py}) and Hamiltonians (\texttt{hamiltonians.py}), and they are used along with the other initial parameters set in the InputParams data class (\texttt{parameters.py}) to set up the simulation (left column on the diagram).
One script (\texttt{simulation.py}) contains the time evolution functions that use the initial parameters to conduct the simulation and obtain the time evolution information of the system state, stored in the Bins data class (\texttt{parameters.py}). These time-evolved states can then be used by declaring operators and taking expectation values (\texttt{operators.py}) or calculating two-time point correlation functions of the output field (\texttt{correlation.py}).

The graph in the right column of the diagram in Fig.~\ref{fig:diagram} shows some of the possible outputs that the user can calculate in a non-Markovian simulation. As we will see in the next sections, these include population dynamics (such as a TLS population or the population of the photon flux at various parts of the waveguide, such as the transmitted flux or the flux in a feedback loop), photon-photon and photon-matter and matter-matter correlations, entanglement, spectra, and more. 

\subsection{Initial states}

As we saw in Sec.~\ref{sec:theory}, in the MPS scheme, the initial state is written as a matrix product state, and only the {\it relevant bins} will be required for the time evolution at each time step. This highly reduces the Hilbert space of each part of the state. 

On the one hand, the system part, containing the quantum emitter information, must be defined. As we will see in the following examples, this depends on the number of emitters and the initial preparation we want them to start with. Since we are working here with TLSs, each emitter will have a bin size of dimension 2, which corresponds to its ground and excited energy levels.
For example, a TLS initially excited can be defined with \texttt{qmps.states.tls\_excited}.

On the other hand, we also need to define the initial field. For example, if we are starting with our waveguide in a vacuum state (\texttt{qmps.states.vacuum}), or if we are sending photons into the waveguide. As we will see in the example in Sec.~\ref{subsec:fock}, with MPS, we have the ability to send quantized photons, e.g., Fock pulses
(\texttt{qmps.states.fock\_pulse}), where we can choose the number of photons, their direction of propagation, and their pulse shape. Since the dimension of each bin determines the maximum number of photons per time bin, for exact Fock state simulation this must be chosen correctly to accommodate probabilities of $N$ photons in a single bin. For example, if we limit the bins to a single photon, the dimension will be two, to contain the zero- and one-photon subspaces. In general, for $N$ photons per time bin, we need an $N+1$ bin size. In addition, if we consider two channels in the waveguide, with photons moving to the right and left, the size of each time bin's dimension will be a product of the individual channel dimensions.

Since this part of the initial state is independent of the system part, it can be implemented for a system containing multiple emitters, even if they are in the non-Markovian regime, making this approach highly powerful.  

This is not the only way to send photons to the system; the drive can also be treated classically, as a Rabi frequency, including it directly in the Hamiltonian. This can be a good approximation in some scenarios when dealing with many photons in the pump field (such as a high-intensity laser field). 

\subsection{Hamiltonian}

In order to evolve our system, as seen in Eq.~\eqref{eq:t_evol}, we need to apply the time evolution operator at each time step. This means writing it in its MPO form (Sec.~\ref{subsec:mps}), and consequently, writing the Hamiltonian in the same form.  

Each Hamiltonian is specific to its system. For example, for a single TLS in an infinite waveguide, the Hamiltonian in the time domain and rotating frame is, 
\begin{equation}
    \begin{split}
     H(t)&= \frac{\Omega(t)}{2} ( \sigma^+ + \sigma^- ) \\
     &+ \sqrt{\gamma_{L}} \left (\sigma^+ b_{L}(t)  +  \sigma^- b_{L}^\dagger(t)\right)
      \\ &+
     \sqrt{\gamma_{R}} \left(\sigma^+ b_{\rm R}(t)  +  \sigma^- b_{R}^\dagger(t)\right),
    \end{split}
\end{equation}
where $\gamma_{L/R}$ are the left/right decay rates, and $\Omega(t)$ can contain a classical pump (which can be a continuous wave pump or pulsed light). This can be written as an MPO, using:
\begin{equation}
    \begin{split}
        &{\cal H}(t_k) = \frac{\Omega(t_k)}{2} \Delta t \left( \sigma^+ +  \sigma^-\right) \\
        &+ \sqrt{\gamma_{L}}\left( \sigma^+ \Delta B_{L}(t_k) + \sigma^- \Delta B_{L}^{\dagger}(t_k) \right)\\
        &+  \sqrt{\gamma_{R}}\left( \sigma^+ \Delta B_{R}(t_k) + \sigma^- \Delta B_{R}^{\dagger}(t_k) \right),     
    \end{split}
\end{equation}
where ${\cal H}(t_k) = H (t_k) \Delta t$ is the Hamiltonian per time step, the quantum operators are now MPOs, and this form is then introduced in the time evolution operator.
In the QwaveMPS package, this case can be simply implemented with the pre-defined function \texttt{qmps.hamiltonian\_1tls}.  

For convenience, more complicated systems are already available in QwaveMPS, including 
\begin{itemize}
    \item Single TLS with a side mirror (semi-infinite waveguide): \texttt{qmps.hamiltonian\_1tls\_feedback}.
    \item Two TLSs in the Markovian regime: \texttt{qmps.hamiltonian\_2tls\_mar}.
    \item Two TLSs in the non-Markovian regime (with delayed time feedback effects): \texttt{qmps.hamiltonian\_2tls\_nmar}.
\end{itemize}

Moreover, these forms can be extended to accommodate other Hamiltonians, including more TLSs or other types of quantum emitters (such as harmonic oscillators
and multi-level systems).

\subsection{Simulation and observables}

Once the initial parameters are defined, we can introduce them as inputs in the main simulation and evolve the system. There are two main types of evolutions, one for systems in the Markovian regime: \texttt{qmps.t\_evol\_mar}, and one for the non-Markovian regime: \texttt{qmps.t\_evol\_nmar}. 

These functions will give us the evolution of both the system and the field bins, which are used afterwards to calculate observables. These observables can easily be calculated with the time evolved bins using functions in the \texttt{operators.py} and \texttt{correlation.py} scripts and passing them the user's operators of interest. Observables include:
\begin{itemize}
    \item Population dynamics. Evaluated using \texttt{qmps.operators.single\_time\_expectation}.
    \item Two-time quantum correlation functions. For example: \texttt{qmps.correlation.correlation\_4op\_2t}. 
    \item Spectra. For example: \texttt{qmps.correlation.spectrum\_w}.
    \item Entanglement entropy: \texttt{qmps.operators.entanglement}.
\end{itemize}

\subsection{Waveguide-QED operations handled automatically}
\label{subsec:automation}

The MPS structure underlying a time-bin waveguide-QED simulation is similar to other one-dimensional chains of local multilevel degrees of freedom, and it can
in principle be constructed with a general-purpose TN library such as {\em ITensor}, {\em mpnum} or {\em MPSDynamics.jl}~\cite{itensor,Suess2017,mpsdynamicsjl_2024}. However, these libraries do not provide the waveguide-QED layer. Left to the user are the waveguide-QED model itself, the time-bin evolution, the delayed interaction together with the swap
sequence that implements it, the gauge (in this case a mixed canonical form) and truncation bookkeeping, the matrix-product factorization of quantum input pulses, and the calculation of field observables and multi-time correlation functions.
Together these account for most of the implementation work in a waveguide-QED MPS calculation, and they recur in every example of Sec.~\ref{sec:examples}.

The central practical contribution of QwaveMPS is that all of these operations are performed internally,
behind a small set of calls whose arguments are largely physical parameters rather than TN objects.
Table~\ref{table:automation} lists them explicitly, alongside the corresponding QwaveMPS function.
Accordingly, a user who understands the physics of the problem, but not the details of tensor networks, can set up,
evolve, and analyze the waveguide-QED systems shown below without writing any MPS code, in the same spirit in which {\em QuTiP} makes master-equation solvers available without requiring the user to implement them.

\begin{DIFnomarkup}
\onecolumngrid
\medskip
\begin{center}
\begin{minipage}{\textwidth}
\captionof{table}{Waveguide-QED-specific TN operations that are performed automatically by QwaveMPS, alongside the corresponding QwaveMPS function.
}
\label{table:automation}
\small
\begin{tblr}{
  colspec = {|Q[t, wd=0.33\textwidth]|Q[t, wd=0.60\textwidth]|},
  hlines,
}
    QwaveMPS Function & Performed Automatically \\
    \texttt{qmps.hamiltonian\_*} & Construction of the per-time-step Hamiltonian MPO from physical parameters (coupling terms $\gamma_L$, $\gamma_R$, drive $\Omega(t)$, detuning $\delta$, mirror phase $\phi$, delay $\tau$), in the 2-site (Markovian) or 3-site (non-Markovian) form.
    \\
    \texttt{qmps.t\_evol\_mar} & SVD and truncation at each time step, tracking and movement of the orthogonality center, swapping of the output-field time bin, and storage of the Schmidt spectra at the system. \\
    \texttt{qmps.t\_evol\_nmar} & Similar to \texttt{qmps.t\_evol\_mar} but now in addition: Identification of the feedback bin at $\tau$, generation and application of the swap network that brings it adjacent to the system and present bins, ${\rm round}(\tau/\Delta t)$ swaps per time step, and restoration of the chain ordering afterwards. \\
    \texttt{bond\_max} (in \texttt{InputParams}) & Bond-dimension control, with an automatic warning raised when the requested cutoff is insufficient to represent the entanglement generated. \\
    \texttt{qmps.fock\_pulse} & Creates an MPS factorization of an $N$-photon wavepacket, accounting for an arbitrary envelope shape and propagation direction, applying the required bin dimensions ($N+1$ per channel) and the tensor product over the left and right channels. \\
    \texttt{qmps.single\_time\_expectation}, \texttt{qmps.b\_pop\_l/r}, \texttt{qmps.b\_pop}, \texttt{qmps.loop\_integrated\_statistics} & Mapping of physical observables (emitter populations, output flux, occupation of the feedback loop) onto the correct bins, including the positional bookkeeping that relates a bin index to a space-time point on the waveguide. \\
    \texttt{qmps.correlation\_*}
    & Application of operators at two separated times with re-gauging in between and assembly of the two-time grid. \\
    \texttt{qmps.spectrum\_w} & Transformation of the steady-state first-order correlation function into the long-time-limit spectrum. \\
    \texttt{qmps.entanglement} & Calculate entanglement entropies at the system and feedback cuts, obtained from the stored Schmidt coefficients. \\
\end{tblr}
\end{minipage}
\end{center}
\medskip
\twocolumngrid
\end{DIFnomarkup}

\section{Selected Simulation Examples}
\label{sec:examples}

In this section, we present selected examples of the QwaveMPS package being used to calculate different waveguide-QED systems and scenarios, and show results for some of the main observables discussed in the previous section. Here, we include computational times of the observables calculated in the examples. 

All the examples below have been run on a single standard computer workstation, specifically with 527.8 GB RAM, 3.00 GHz, and 18 cores, though {only a maximum of around 180~MB ($<0.2$~GB) of RAM was necessary in the process used to compute even the most complicated examples considered (including significant delay times, non-Markovian dynamics, multi-time correlation functions, and nonlinear effects)}. Thus, all of the simulations and examples shown below {\it can easily be run even on a modest laptop computer}. Although they can also be run on HPC resources, at this time, GPU acceleration is not supported.

\subsection{Vacuum population dynamics in the linear (one quantum) regime}
\label{subsec:linear_regime}
\begin{figure}[ht]
  \centering
  \includegraphics[width=\columnwidth]{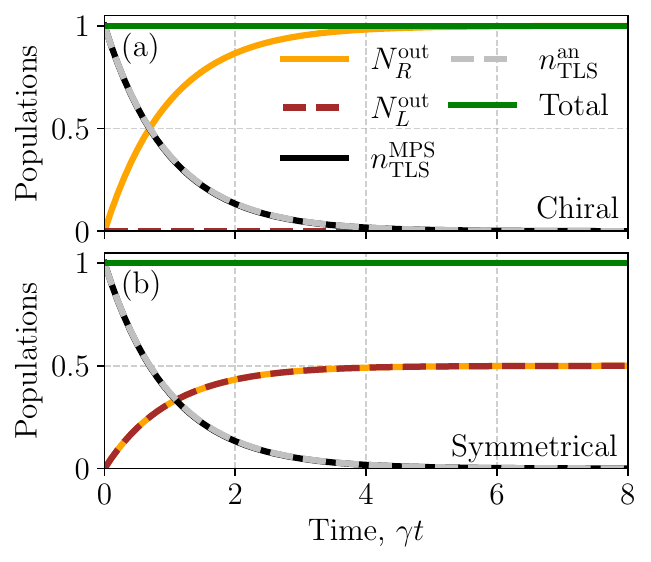}
  \caption{Decay of a single excited TLS in an infinite waveguide, with (a) right chiral coupling, and (b) symmetrical coupling to the waveguide. The TLS population dynamics is calculated with MPS (black) and benchmarked with the analytical solution (dashed grey); we also show the integrated photon flux for right (orange) and left (dashed brown) channels, and a quanta conservation check (green). }
  \label{fig:1tls_m}
\end{figure}

We begin by calculating a very simple case, which has a well-known analytical solution, to show how the package works and check the (known) solution. This example corresponds to the decay of a single TLS in an infinite waveguide (Fig.~\ref{fig:1tls_m}). As we saw in the previous section, we need to set up the initial parameters, including the type of coupling and the initial state, as shown below:
\begin{python}
""" Initial state and coupling"""
gamma_l,gamma_r=
qmps.coupling('symmetrical',gamma=1)
input_params=qmps.InputParams(
delta_t=0.05, #time step
tmax=8, #max time
d_sys_total=np.array([2]), #TLS bin size
d_t_total=np.array([2,2]),#time bin size 
gamma_l=gamma_l, #left coupling
gamma_r=gamma_r, #right coupling 
bond_max=4 #max bond dimension
)
i_s0=qmps.tls_excited()
i_n0=qmps.vacuum(tmax,input_params)
\end{python}

Here, we have chosen an initially excited TLS ($i_{s0}$), with the waveguide initially in vacuum, and symmetrical coupling. As mentioned in the previous section, since we have symmetrical coupling, each time bin accounts for both the left and right channels, in this case, limiting them to up to 1 photon per time step in each channel. The maximum bond dimension remains low since we only have one excitation.
We also need to choose the Hamiltonian:
\begin{python}
"""Hamiltonian"""
Hm=qmps.hamiltonian_1tls(input_params)    
\end{python}
which in this case corresponds to the 1-TLS Hamiltonian.

Once we have the initial parameters defined, the system is evolved in time to calculate the TLS bins, field bins, and the Schmidt coefficients, which can be seen as explicit constituents of the Bins class:
\begin{python}
"""Time evolution of the system"""
bins=qmps.t_evol_mar
    (Hm,i_s0,i_n0,input_params)
sys_states=bins.system_states
field_states=bins.output_field_states
\end{python}

We can then define the relevant operators that we will use to take expectation values:
\begin{python}
"""Define Operators of interest"""
tls_pop_op=qmps.tls_pop()
flux_op_l=qmps.b_pop_l(input_params)
flux_op_r=qmps.b_pop_r(input_params)
flux_ops=[flux_op_l,flux_op_r]
\end{python}
and using the bins and these operators we can calculate different observables, such as the population dynamics:
\begin{python}
"""Calculate population dynamics"""
tls_pops=qmps.single_time_expectation
    (sys_states,qmps.tls_pop())

fluxes=qmps.single_time_expectation
    (field_states, flux_ops)
\end{python}

In Fig.~\ref{fig:1tls_m}, we show some of these observables both for chiral [Fig.~\ref{fig:1tls_m}~(a)] and symmetrical (TLS) coupling  [Fig.~\ref{fig:1tls_m}~(b)]. 
First, with the black line, we show the TLS population ($n_{\rm TLS}^{\rm MPS}$) which is benchmarked with the known analytical result of the decay: $n_{\rm TLS}^{\rm an} = \exp(-\gamma t)$, represented with the grey dashed line. We also show the
(time) integrated photon flux:
\begin{equation}
    N^{\rm out}_{\alpha} (t)= \int_0^t n^{\rm out}_{\alpha}(t') dt',
    \label{eq:N_out}
\end{equation}
where $n^{\rm out}_{\alpha}$ corresponds to the flux of photons going towards the right or left in the waveguide (with a scaling of $\gamma$ in our normalized units).
It can be observed that in the symmetrical coupling, both channels behave similarly, where the right integrated flux $N^{\rm out}_{R}$ is shown in orange, and the left one $N^{\rm out}_{L}$ is shown in dashed brown. However, in the chiral solution, where the TLS is only coupled to the right channel, the left integrated flux remains zero at all times, as expected. 

Finally, we have also checked that the total system quanta are conserved, with the following conservation rule:
\begin{equation}
    N_{\rm total}(t) = n_{\rm TLS}(t) +  N^{\rm out}_{R} +  N^{\rm out}_{L},
\end{equation}
which in this case should correspond to one quantum at all times ($N_{\rm total}(t) =1$), since we have only one excitation in the system. This is shown and confirmed on the graph with the solid green curve.

\begin{figure}[ht]
  \centering
  \includegraphics[width=\columnwidth]{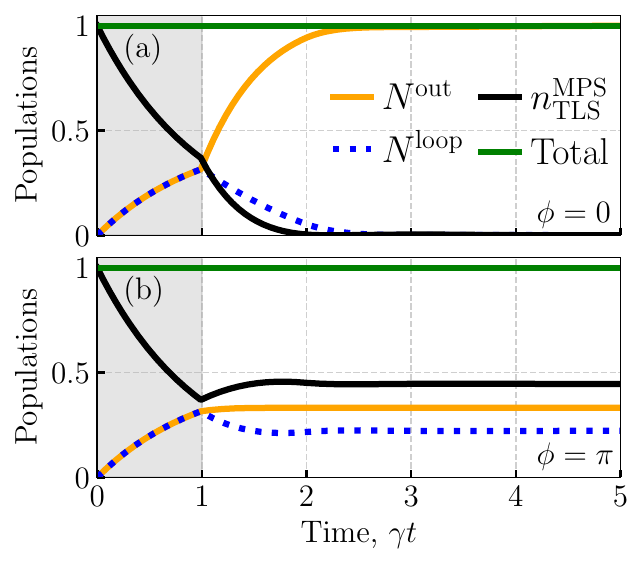}
  \caption{Decay of a single excited symmetrical TLS in a semi-infinite waveguide for a delay time of $\gamma \tau = 1$, with (a) a destructive feedback phase ($\phi=0$), and (b) a constructive feedback phase ($\phi=\pi$). TLS population dynamics (black), integrated photon flux going out of the loop [orange, Eq.~\eqref{eq:N_out}], mean photon number within the loop [dotted blue, Eq.~\eqref{eq:N_in}], and quanta conservation check (green). Grey shaded area represents the time until feedback, from $t=0$ to $t=\tau$.
   }
  \label{fig:1tls_nm}
\end{figure}

After this simplest Markovian case is tested and understood, we can now introduce more complicated feedback effects. 
In practical systems, this can be done, for example, by adding a mirror to one side of the waveguide. In this case, we will have a single TLS in a semi-infinite waveguide (Fig.~\ref{fig:1tls_nm}). 
The initial parameters are set up in a very similar manner, but we now have to consider the finite delay time $\tau= 2d/v_g$, which relates to the distance between the TLS and the mirror, $d$, and the group velocity $v_g$ (of the waveguide mode), and the mirror phase $\phi$. We start with a similar initial state, but now with the corresponding Hamiltonian:
\begin{python}
"""Choose the Hamiltonian"""
Hm=qmps.hamiltonian_1tls_feedback
  (input_params)
\end{python}

In this case, the photons decay from the TLS in both directions (unless we had a chiral system), and the left-moving photons will travel, hit the mirror, reflect back to the right, and interact again with the TLS. We enter here the non-Markovian regime, and the corresponding time evolution of the system is: 
\begin{python}
""" Time evolution of the system"""
bins=qmps.t_evol_nmar(Hm,i_s0,i_n0,
    input_params)
\end{python}

Due to the symmetry of this feedback loop, in this case, we have only one channel for the field, and we can calculate not only the transmitted photons to the right with the integrated photon flux, but also the mean number of photons {\it within the loop}, which can be calculated as:
\begin{equation}
    N^{\rm loop}(t)= \int_{t-\tau}^t n^{\rm loop}(t') dt',
    \label{eq:N_in}
\end{equation}
where $n^{\rm loop}$ is the photon flux inside the loop. We stress that this is a significant difference from Markov simulations, which would not allow any photons in the loop.

Similarly to before, we can obtain the population dynamics, now in the loop as well, by acting on the appropriate states and using the single-channel operators:
\begin{python}
""" Calculate population dynamics"""
flux_op=qmps.b_pop(input_params)
loop_states=bins.loop_field_states
loop_flux=qmps.single_time_expectation
    (loop_states,flux_op)

loop_p=qmps.loop_integrated_statistics
    (loop_flux,input_params)
\end{python}

Here we have calculated the flux into the loop, and have used its result with the helper function \texttt{loop\_integrated\_statistics} to obtain the mean number of photons within the loop.

In this time-delayed feedback example, we show population dynamics for a delay time of $\gamma \tau=1$, and for a destructive and a constructive phase, in Figs.~\ref{fig:1tls_nm}(a) and \ref{fig:1tls_nm}(b), respectively. First, in Fig.~\ref{fig:1tls_nm}(a), we choose a phase $\phi=0$, which will lead to a destructive interaction between the feedback photons and the TLS. This is observed in the TLS population dynamics (black), where there is a decay similar to the one observed in Fig.~\ref{fig:1tls_m}, until the feedback time $t=\tau$ ($\gamma t = 1$), and then there is an increased decay, and an increase of the integrated flux (orange). In addition, we also show the mean number of photons within the loop (dotted blue), which in this case grows as the TLS decays, and then decreases again after the feedback time. 

In Fig.~\ref{fig:1tls_nm}(b), we choose a constructive phase of $\phi=\pi$. The dynamics before the feedback time remain the same, but after $\gamma t =1$, we observe completely different dynamics. Now the photons get {\it trapped} in the feedback loop, and there is a long-lasting steady state. This is shown with finite constant values of the calculated observables. 

\begin{figure}[ht]
  \centering
  \includegraphics[width=\columnwidth]{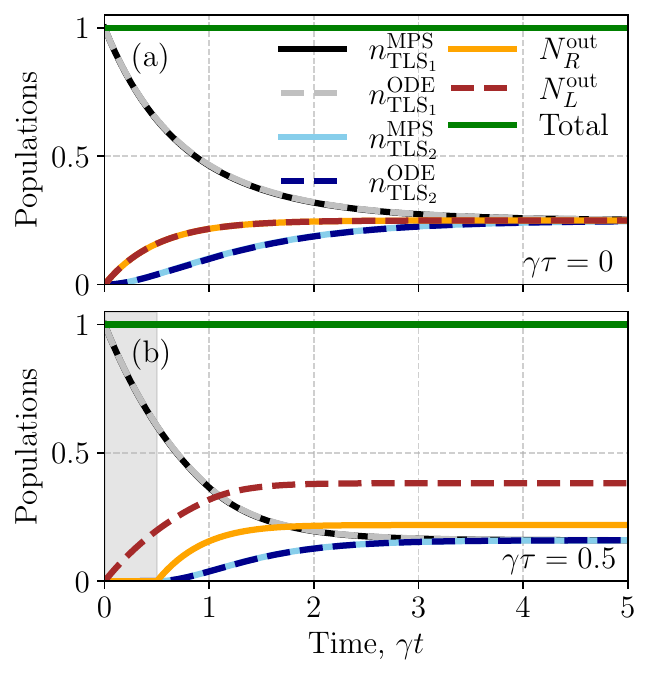}
  \caption{Two TLSs symmetrically coupled to an infinite waveguide. The first TLS is initially excited, the second one starts in the ground state, and the phase between atoms is $\phi=\pi$. (a) Markovian regime ($\gamma \tau = 0$), and (b) non-Markovian regime ($\gamma \tau = 0.5$). Emitter populations (black for TLS 1 and cyan for TLS 2), which are benchmarked with a separate ODE solver (dashed grey and dark blue, respectively).
  Excellent agreement is shown. Integrated right flux (orange) and left one (dashed brown), and total quanta check (green).}
  \label{fig:2tls_lin}
\end{figure}

The final example shown here of vacuum dynamics in the linear regime corresponds to the addition of a second TLS (Fig.~\ref{fig:2tls_lin}). In this case, the initial state will be coded in the TLSs part as an outer product of the tensor spaces:
\begin{python}
""" Initial state and coupling"""
i_s01=qmps.tls_excited()
i_s02= qmps.tls_ground()
i_s0=np.kron(i_s01,i_s02)

gamma_l1,gamma_r1=qmps.coupling
    ('symmetrical',gamma=1)
gamma_l2,gamma_r2=qmps.coupling
    ('symmetrical',gamma=1)
\end{python}
where we have chosen the first (left) TLS in an initial excited state and the second one in the ground state, and symmetrical coupling for both TLSs. It is important to note here that, although we have chosen this particular initial state, we could easily start with an initially entangled state:
\begin{python}
i_s01=qmps.tls_excited()
i_s02=qmps.tls_ground()

i_s0=1/np.sqrt(2)*(np.kron(i_s01,i_s02)
+np.kron(i_s02,i_s01))    
\end{python}
or any other combination.

We can again study the Markovian and the non-Markovian regimes. In the first case, the Hamiltonian is
\begin{python}
Hm=qmps.hamiltonian_2tls_mar
    (input_params)
\end{python}
and the time evolution function is general for any number of quantum emitters in the Markovian regime. Hence, the time bins are calculated in the same manner as in the previous case. The population observables can be calculated in the same way as before, except that individual system operators must be extended to act on the entire system state space:
\begin{python}
tls1_pop_op=np.kron(tls_pop_op,d_sys2)
tls2_pop_op=np.kron(d_sys1,tls_pop_op)
tls_ops=[tls1_pop_op,tls2_pop_op]

tls_pops=qmps.single_time_expectation
    (sys_states,tls_ops)
\end{python}
which is calculated identically in the Markovian and non-Markovian cases, though when we introduce feedback, we need to use the non-Markovian version of the Hamiltonian:
\begin{python}
Hm=qmps.hamiltonian_2tls_nmar
    (input_params)
\end{python}
Again, the time evolution function in the non-Markovian regime is general for any number of quantum emitters, with a single non-Markovian loop for now, and the bins have been calculated as before.

\begin{figure*}[ht]
  \centering
  \includegraphics[width=0.85\textwidth]{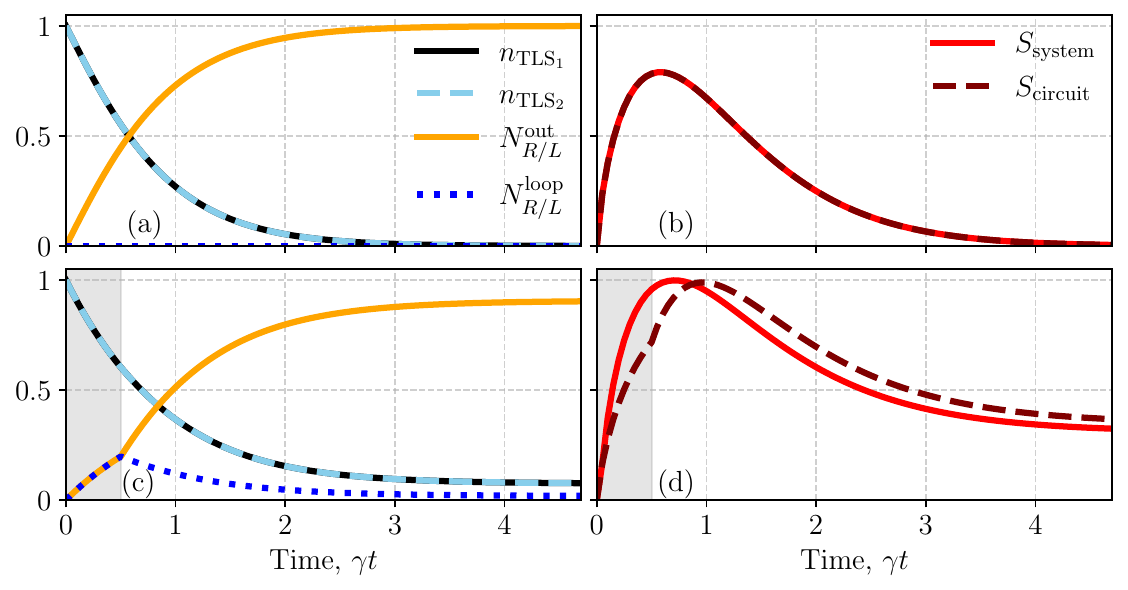}
  \caption{Two TLSs in an infinite waveguide. Both TLSs start excited, and the phase between atoms is $\phi=0$. (a,b) Markovian regime ($\gamma \tau = 0$), and (c,d) non-Markovian regime ($\gamma \tau = 0.5$), where the shaded area represents time until $\tau$. (a,c) Left and right TLS population dynamics are shown in black and dashed cyan, respectively. Integrated output flux, $N^{\rm out}_{R/L}$ in orange and photons moving in the waveguide between TLSs, $N^{\rm loop}_{R/L}$ in dotted blue. (b,d) Entanglement entropy between the TLSs and the waveguide field, $S_{\rm system}$ in red, entropy between the TLSs plus the section between them, and the rest of the waveguide field in dashed maroon. 
  }
  \label{fig:2_TLS_nonlin}
\end{figure*}

Figure~\ref{fig:2tls_lin} shows example population dynamics in the Markovian regime in (a) and the non-Markovian regime in (b), with a feedback time of $\gamma \tau=0.5$. Both cases are calculated with a phase of $\phi=\pi$ between the TLSs, and with the left TLS initially excited. The TLS population of the left TLS (black) and the right one (cyan) have been benchmarked with an ordinary differential equations (ODE) solver for the following equations,
\begin{equation}
\begin{split}
     &\frac{d \braket{\sigma_i^+(t)\sigma_i^-(t)}}{dt}  = -\frac{\gamma}{2} \Big[ \braket{\sigma_i^+(t)\sigma_i^-(t)}\\ 
      &+\braket{\sigma_j^+(t-\tau)\sigma_j^-(t-\tau)} e^{i\phi} \theta(t-\tau) \Big],       
      \label{eq:ODE}
\end{split}
\end{equation}
where $i=1,2$ and $j=1,2$ with $j \neq i$ for each TLS, $\gamma/2 = \gamma_{R} = \gamma_{L}$ for symmetrical coupling, $\tau$ is the delay time, $\phi$ is the phase between the TLSs, and $\theta(t-\tau)$ is the Heaviside step function~\cite{Liu2025}. 

To solve Eq.~\eqref{eq:ODE}, we utilize the SciPy-based Delay Differential Equation 
solver \texttt{ddeint}, which is designed to efficiently solve time-delayed coupled ODEs in Python.
The ODE solutions are shown in dashed grey and dark blue for the first and second TLS, respectively, and perfect agreement is achieved between both approaches. It is important to note here that even though this Python package is highly optimized, run times are comparable, on the order of a second, and we stress that Eq.~\eqref{eq:ODE} is limited to the single excitation case (linear regime, which is a {\it severe restriction}, as the uniquely quantum dynamics are beyond weak excitation). 

The differences between the Markovian limit in (a) and the feedback effects in (b) are significant in all observables. In the case of the TLSs dynamics, we can see how the right TLS starts getting populated from $\gamma t=0$ in (a), but it will remain zero in (b) until $t=\tau$. In addition, the symmetry of the integrated flux is broken when considering feedback, with a lower $N^{\rm out}_{R}$ and a higher $N^{\rm out}_{L}$ in the non-Markovian results. Finally, both cases show a quanta conservation check (green) which must remain at one since we have only one excitation (linear regime).

In Table~\ref{table:linear_vacuum}, we show average run times when calculating the population dynamics of the previous examples. In all cases, the calculations
take less than a second, showing the power and high efficiency of the MPS method.
The run times for one and two emitters are the same to within the scatter between
repeated runs.

\begin{table}[h]
\centering
\caption{Run times, using a single computer workstation (specific hardware specs are given in the main text), for population dynamics of the examples presented in Sec.~\ref{subsec:linear_regime} (vacuum dynamics) in the Markovian (`M') and non-Markovian (`NM') regimes. Parameters: time step $\gamma \Delta t = 0.05$, final time $\gamma t_{\rm max}=8$, and maximum bond dimension $\chi_{\rm max} = 4$. 
}
\begin{tabular}{|c|c|c|}
    \hline
    System & Delay time ($\gamma \tau$) & Run time \\
    \hline
    1 TLS M & 0 & 0.08 s \\
    1 TLS NM & 1 & 0.68 s \\
    2 TLS M & 0 & 0.06 s \\
    2 TLS NM & 0.5 & 0.67 s \\
    \hline
\end{tabular}
\label{table:linear_vacuum}
\end{table}

\subsection{Vacuum population dynamics and entanglement entropy in the 
(two quanta) nonlinear regime}
\label{subsec:vac_nonlinear}

So far, we have explored only linear dynamics, 
which is common in waveguide-QED studies as it is easier to model and understand (otherwise, a Markov approximation is usually invoked, which is not suitable for modeling strong correlations and entanglement between photons and matter). However, the real power of MPS is that it is 
not limited to single excitations, and can easily go to two, three, four, and more. 

In this next example, we study the same 2-TLS system as the one shown in Sec.~\ref{subsec:linear_regime}, but now we start the system with both TLSs excited (two quanta, which is intrinsically nonlinear). The initial state is:
\begin{python}
i_s01=qmps.tls_excited()
i_s02=qmps.tls_excited()

i_s0=np.kron(i_s01,i_s02)

i_n0=qmps.vacuum
    (tmax,input_params)    
\end{python}

The Hamiltonian, time evolution and population functions used are the same as in Sec.~\ref{subsec:linear_regime}. Additionally, in this example, we introduce a new observable: entanglement entropy. First, the entanglement entropy is defined between two parts of a system. In this case, we show the entanglement entropy between the TLSs and the waveguide~\cite{PhysRevResearch.3.023030,Guimond_2017}:
\begin{equation}
    S_{\rm system}=-\sum_\beta \Lambda[{\rm sys}]^2_\beta  \log_2(\Lambda[{\rm sys}]^2_\beta),
    \label{eq:ent_system}
\end{equation}
where $\Lambda[{\rm sys}]_\beta$ are the Schmidt coefficients corresponding to the TLSs system bins and $\beta$ represents the position of these coefficients. 

In the non-Markovian regime, we also show the entanglement entropy between the `circuit' formed by the 2 TLSs plus the waveguide section that connects them, and the rest of the waveguide (in the Markovian limit, both are the same since the space between the TLSs is not considered). This follows an equation similar to Eq.~\eqref{eq:ent_system} but with the Schmidt coefficients of the feedback bins $\Lambda[\tau]_\beta$~\cite{slzl-d5dz}:
\begin{equation}
S_{\rm circuit}=-\sum_\beta \Lambda[\tau]^2_\beta \log_2(\Lambda[\tau]^2_\beta).  
\end{equation}
This is implemented with the following functions:
\begin{python}
#entanglement entropy
ent_s=qmps.entanglement(bins.schmidt)
ent_tau=qmps.entanglement
    (bins.schmidt_tau)
\end{python}

In Fig.~\ref{fig:2_TLS_nonlin}, both Markovian (a,b) and non-Markovian dynamics (c,d) for a delay time of $\gamma \tau =0.5$ are shown. In this case, the phase between the TLSs is $\phi=0$. Here, the left column plots (a,c) again show population dynamics with the decay of TLS 1 (black) and TLS 2 (dashed cyan), and the corresponding integrated fluxes $N^{\rm out}_{R/L}$ (orange). In addition, the mean number of photons between the TLSs is calculated, $N^{\rm loop}_{R/L}$ (dotted blue). We recognize how, in the non-Markovian regime, there is a small trapping of the population in the TLSs as well as in the section of the waveguide between them.

The right column panels, i.e., Figs.~\ref{fig:2_TLS_nonlin}(b) and
\ref{fig:2_TLS_nonlin}(d), show the entanglement entropy. In the Markovian regime $S_{\rm system} = S_{\rm circuit}$ as there is no waveguide section considered between TLSs, but as feedback effects appear, $S_{\rm system}$ (red) and $S_{\rm circuit}$ (dashed maroon) deviate from each other. In addition, in this second case, both entropies reach the maximum of one, which is not the case in the Markov limit (b).
This example study shows very interesting nonlinear dynamics that are challenging to calculate using other theoretical and modeling approaches.   

Table~\ref{table:nonlinear_vacuum} shows the run times for the populations, and the entanglement entropy of the example shown here, both in the Markovian limit, and in the non-Markovian regime with a delay time of $\gamma \tau=0.5$. Even though we are in the nonlinear regime, all the results were calculated in less than a second, which again demonstrates the strength and efficiency of this MPS approach. In this case, we chose a maximum bond dimension of 8 to resolve the dynamics, since we now have two excitations involved and non-Markovian dynamics.  

\begin{table}[h]
\centering
\caption{Run times (single computer) for the observables calculated in Sec.~\ref{subsec:vac_nonlinear} (nonlinear vacuum dynamics) in the Markovian and non-Markovian regimes. Parameters: time step $\gamma \Delta t = 0.05$, final time $\gamma t_{\rm max}=5$, and maximum bond dimension $\chi_{\rm max} = 8$.
}
\begin{tabular}{|c|c|c|}
    \hline
    Observable & Delay time ($\gamma \tau$) & Run time \\
    \hline
    Population & 0 & 0.06 s \\
    Entropy & 0 & 0.01 s \\
    Population & 0.5 & 0.70 s \\
    Entropy & 0.5 & 0.02 s \\
    \hline
\end{tabular}
\label{table:nonlinear_vacuum}
\end{table}

\subsection{Two-level system driven by a classical pump field with and without time-delayed feedback, in the nonlinear regime: correlations and feedback-modified Mollow spectra}
\label{subsec:classical_pump}

In the next examples, we study the behavior of a single TLS when it is driven by a continuous-wave (CW) laser, and then by a time-dependent (classical pulse) pump.

\begin{figure*}[t]
  \centering
  \includegraphics[width=\textwidth]{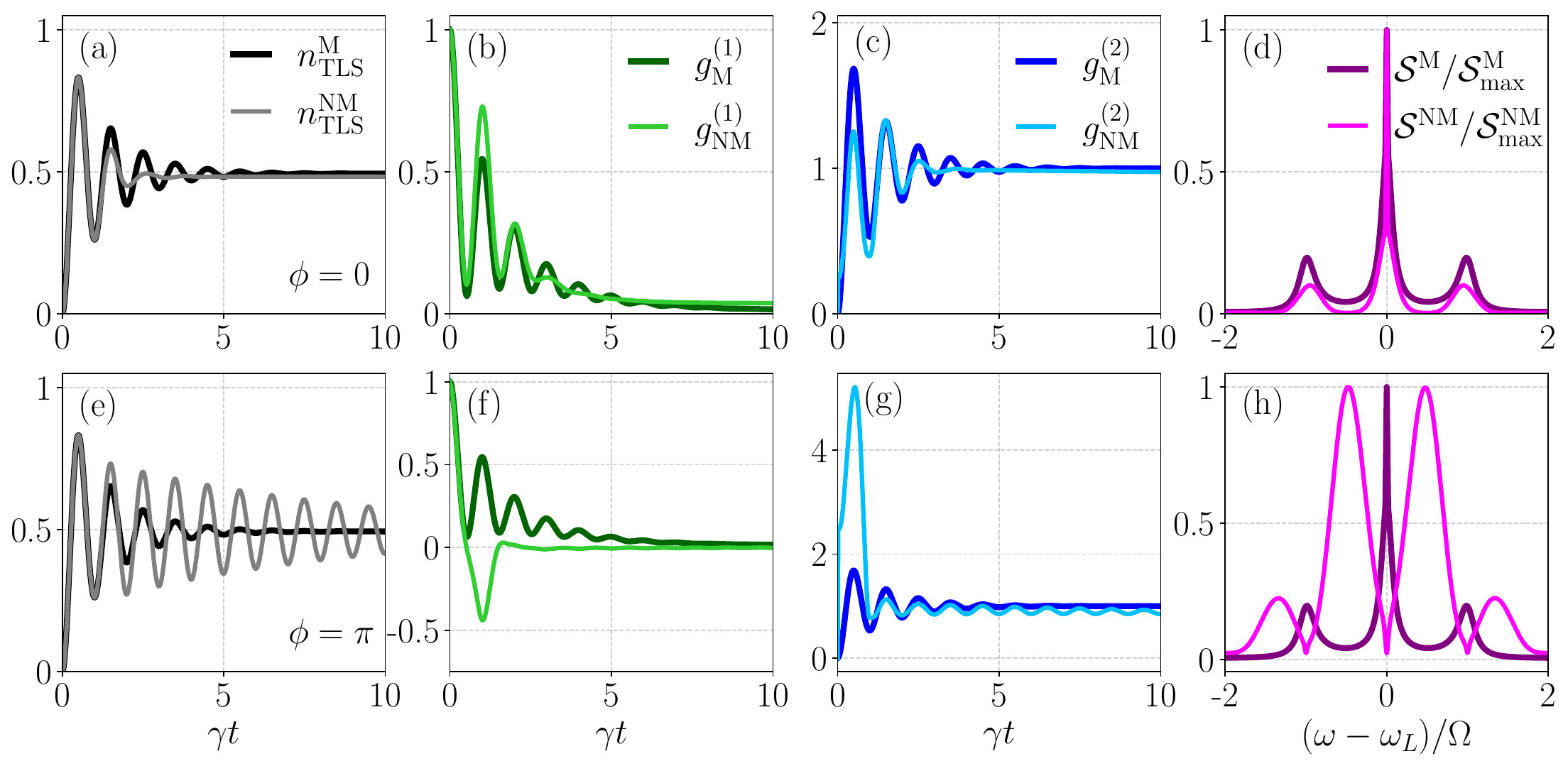}
  \caption{Single TLS under a classical CW pump field in the Markovian limit (labeled with $M$), in an infinite waveguide, and with non-Markovian effects, in a semi-infinite waveguide (labeled with $NM$) with a delay time $\gamma \tau =1$, where (a,b,c,d) have a phase $\phi=0$ and (e,f,g,h) a phase $\phi=\pi$. The pump strength is $\Omega=2\pi\gamma$ in all cases. (a,e) Population dynamics, (b,f) steady-state normalized first-order correlation functions, (c,g) steady-state normalized second-order correlation functions, and (d,h) long-time limit spectra, $\mathcal{S}$ normalized to its maximum value.
  }
  \label{fig:classical_pump}
\end{figure*}

\subsubsection{Continuous-wave classical pump field}
\label{subsubsec:cw_pump}

We start by pumping the TLS vertically with a classical CW pump field (Fig.~\ref{fig:classical_pump}), hence, the external field does not propagate through the waveguide~\cite{Grimsmo2015,PhysRevResearch.3.023030}. This is a challenging calculation since, as the loop length increases, a higher number of photons in the loop needs to be accounted for in order to get the correct results, and theories limited to a low number of photons start failing. 
Note that one could also excite the TLS via the waveguide with a coherent pump, which in the single TLS case can be performed with an input-output relationship, not adding difficulty to the problem.  

In this case, we choose the TLS initially in the ground state, and the waveguide in vacuum. We also choose the strength of the pump; in this example, we have chosen a Rabi frequency of $\Omega=2\pi\gamma$, and then it is added to the total Hamiltonian as follows:
\begin{python}
cw_pump=2*np.pi

Hm=qmps.hamiltonian_1tls
    (input_params,cw_pump)
\end{python}
It is important to note that the code is in units of $\gamma$. Note also that this is a solution for a resonant drive, but if we want to pump the system off-resonance, we can add the detuning \texttt{delta} as an additional parameter (which is zero by default):
\begin{python}
cw_pump=2*np.pi
delta= np.pi

Hm=qmps.hamiltonian_1tls
    (input_params,cw_pump,delta)
\end{python}

Then, we can calculate the time evolution using the same functions as in the previous section, and calculate population dynamics in a similar manner. 

Here, we also introduce the two-time quantum correlation functions,
\begin{equation}
    G^{(1)}_{\alpha \alpha'}(t,t+t')= \braket{b^{\dagger}_{\alpha}(t) b_{\alpha'}(t+t')}
    \label{eq:G1_twotimes}
\end{equation}
and
\begin{equation}
\begin{split}
    &G^{(2)}_{\alpha \alpha'}(t,t+t')= \\
    &\braket{b^{\dagger}_{\alpha}(t) b^{\dagger}_{\alpha'}(t+t')b_{\alpha'}(t+t') b_{\alpha}(t)},
    \label{eq:G2_twotimes}
\end{split}
\end{equation}
where we use the variable $t'$, rather than the more usual $\tau$,
since we have used $\tau$ earlier for the definition of time delay in the feedback systems.

In the case of working with a right and left waveguide channel (most of the cases of interest), we can calculate the correlation of photons traveling to the right ($G_{RR}$), traveling to the left ($G_{LL}$), or correlations between right and left photons and vice versa ($G_{RL}$ and $G_{LR}$). 

As it is already known in the literature, the dynamics of a CW driven TLS reach a steady state, allowing one to calculate two-time correlation functions once the system has reached this state, with
\begin{equation}
    G^{(1)}_{\alpha \alpha'}(t_{ss},t_{ss}+t')= \braket{b^{\dagger}_{\alpha}(t_{ss}) b_{\alpha'}(t_{ss}+t')}
\end{equation}
and
\begin{equation}
\begin{split}
    &G^{(2)}_{\alpha \alpha'}(t_{ss},t_{ss}+t')= \\
    &\braket{b^{\dagger}_{\alpha}(t_{ss}) b^{\dagger}_{\alpha'}(t_{ss}+t')b_{\alpha'}(t_{ss}+t') b_{\alpha}(t_{ss})},
\end{split}
\end{equation}
and their normalized versions,
\begin{equation}
   g^{(1)}_{\alpha \alpha'}(t_{ss},t_{ss}+t')=\frac{G^{(1)}_{\alpha \alpha'}(t_{ss},t_{ss}+t')}{\braket{b^{\dagger}_{\alpha}(t_{ss}) b_{\alpha}(t_{ss})}}, 
\end{equation}
and
\begin{equation}
    g^{(2)}_{\alpha \alpha'}(t_{ss},t_{ss}+t')=\frac{G^{(2)}_{\alpha \alpha'}(t_{ss},t_{ss}+t')}{\braket{b^{\dagger}_{\alpha}(t_{ss}) b_{\alpha}(t_{ss})}^2}.
\end{equation}

We have implemented an efficient function to calculate faster two-time correlations in this regime. The interface will be familiar to users of {\em QuTiP}~\cite{johansson_qutip_2013}, and in the two-operator case takes the form
\begin{python}
corr_bins = bins.correlation_bins
correls, t_primes, t_ss=
    qmps.correlation_ss_2op
    (corr_bins,field_states,
    a_ops,b_ops,input_params)
\end{python}
which gives us correlations of the output field for a desired list of operators of the form $\braket{A(t_{ss})B(t_{ss}+t')}$. This function also yields a corresponding list of $t'$ times, as well as the time (in units of $\gamma^{-1}$) at which the steady state is reached. 

An analogous function can be called to calculate four-operator steady state correlations of the form $\braket{A(t_{ss})B(t_{ss}+t')C(t_{ss}+t')D(t_{ss})}$, and if fluxes have already been calculated the normalized correlation functions can also be easily calculated via
\begin{python}
index=int(round(t_ss/delta_t))
g1=correlation_G1/flux[index]    
g2=correlation_G2/flux[index]**2    
\end{python}

Support also exists for generalized two-time steady-state correlations of the output field.

Furthermore, with the solution of the steady state first-order correlation function, we can directly calculate the long-time limit spectrum, using:
\begin{equation}
\begin{split}
    &\mathcal{S}(\omega) = \\
    &{\rm Re} \left[ \int_0^\infty \braket{b^{\dagger}_{\alpha}(t_{ss}) b_{\alpha}(t_{ss}+t')} e^{i(\omega-\omega_{L}) t'} dt' \right],
\end{split}
\end{equation}
where $\omega_{L}$ is the center frequency of the drive (laser, in the case of optical waveguide systems), and $\alpha$ corresponds to the output direction, which can be right or left ($R/L$).  This is implemented as a function; for example, to calculate a right output spectrum,
\begin{python}
spect,w_list=qmps.spectrum_w
    (delta_t,correlation_G1r)    
\end{python}

Figure~\ref{fig:classical_pump} shows a summary of these new observables calculated for a single TLS in an infinite waveguide (Markovian regime) and in a semi-infinite waveguide (non-Markovian case) with a delay time $\gamma \tau= 1$ and mirror phases of $\phi=0$ and $\phi=\pi$. 

In Fig.~\ref{fig:classical_pump}(a), the population dynamics of the Markovian solution and the non-Markovian with a phase $\phi=0$ are shown. Both cases present the characteristic Rabi oscillations of a strongly pumped system, with a partial suppression of these oscillations in the non-Markovian solution (in grey) due to the destructive feedback phase. In (b), the two-time first-order correlation function in the steady-state regime is shown. Here, we again see how, although at the beginning the oscillations seem larger in the non-Markovian solution, they die out faster than in the infinite waveguide case. This is again seen in (c) in the second-order correlation function, where we see oscillations until it reaches an expected value of one at long times.

Finally, Fig.~\ref{fig:classical_pump}(d) shows the long-time limit spectra of both cases, where we can see the Mollow triplet in both solutions, with an enhancement of the peaks in the non-Markovian solution, due to the delayed feedback effects. 

Similar results are shown in panels (e,f,g,h) of Fig.~\ref{fig:classical_pump}, but now comparing the Markovian regime with the non-Markovian regime using a constructive phase $\phi=\pi$. In this case, we can observe population dynamics presenting Rabi oscillations for a longer period of time, showing bunching in the correlations, and blocking the central spectral peak. In addition, we can see new resonances appearing in the non-Markovian spectrum, consistent with the results in~\cite{PhysRevA.106.013714};
we also highlight that similar features have recently been experimentally measured using a superconducting QED system~\cite{chen2026resonancefluorescenceartificialatom}.

Run times are shown in Table~\ref{table:cw_pump}, {\it which show impressive run times, seconds or less, in all simulation examples}. Here, we also highlight that we are dealing with a highly nonlinear quantum system, which requires a higher maximum bond dimension ($\chi_{\rm max} = 18$). The system is now run for longer times $\gamma t_{\max}=40$, in order to get to the steady state and run long enough for the steady state correlations ($\gamma t_{\rm ss} \approx 30$) and the long-time limit spectra. However, calculations remain remarkably fast, with the non-Markovian population dynamics being the only case to go above a single second ($13.2$ seconds). Note that if, in any case, the maximum bond dimension chosen is not large enough, the code will raise a warning for it to be increased.

\begin{table}[h]
\centering
\caption{Run times (single computer) for the observables calculated in Sec.~\ref{subsubsec:cw_pump} where a classical pump is introduced. Parameters: time step $\gamma \Delta t = 0.05$, final time $\gamma t_{\rm max}=40$, and maximum bond dimension $\chi_{\rm max} = 18$. The correlation time from reaching steady state (SS) is $\gamma t_{\rm ss} \approx 30$. 
}
\begin{tabular}{|c|c|c|}
    \hline
    Observable & Delay time ($\gamma \tau$) & Run time \\
    \hline
    Population & 0 & 0.41 s \\
    SS correl.+$\mathcal{S}$ & 0 & 0.70 s \\
    Population & 1 & 13.20 s \\
    SS correl.+$\mathcal{S}$ & 1 & 0.62 s \\
    \hline
\end{tabular}
\label{table:cw_pump}
\end{table}

\subsubsection{Time-dependent classical pulsed pump field}

\begin{figure*}[ht]
  \centering
  \includegraphics[width=0.95\textwidth]{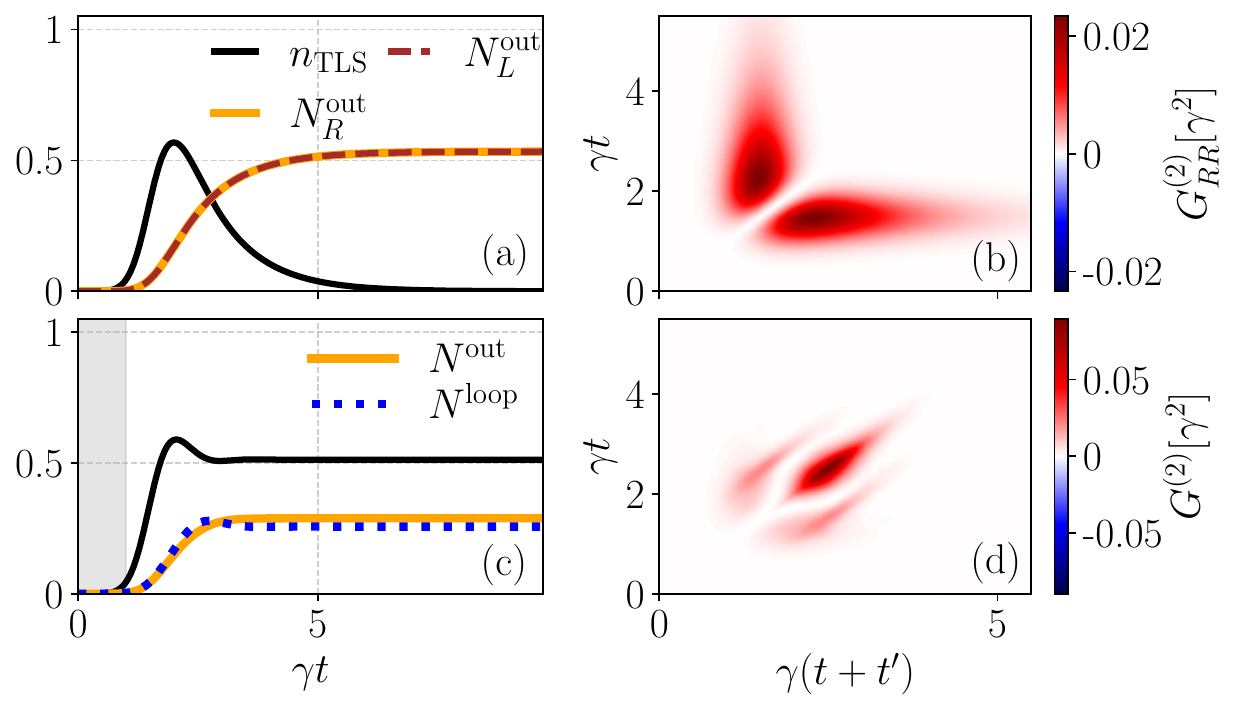}
  \caption{Single TLS driven by a classical $\pi$-pulse, with a width of $\gamma\sigma=0.5$ and centered at $\gamma t_c=1.5$. (a,b) Markovian solution with an infinite waveguide. 
  (c,d) Non-Markovian regime with a semi-infinite waveguide, with a delay time of $\gamma \tau = 1$ and a constructive phase $\phi=\pi$. Population dynamics are shown in (a,c), with (a) TLS population dynamics (black), integrated right (orange) and left (dashed brown) output flux, $N^{\rm out}_R$ and $N^{\rm out}_L$ respectively [Eq.~\eqref{eq:N_out}], and (c) TLS population dynamics (black), output flux (orange) $N^{\rm out}$ [Eq.~\eqref{eq:N_out}] and photons in the feedback loop (dotted blue) $N^{\rm loop}$ [Eq.~\eqref{eq:N_in}]. (b,d) show the second-order correlation functions, with right-right photon scattering $G^{(2)}_{RR}$ in (b) [see Eq.~\eqref{eq:G2_twotimes}], and the single-channel second-order correlation $G^{(2)}$ in (d), where there is only one output.}
  \label{fig:classical_pipulse}
\end{figure*}

The time-discretized nature of MPS allows one to easily extend the previous case to time-dependent pump fields, e.g., the addition of classical pulsed light, where we can choose the temporal shape and the strength of the pulse that we want to use. 

This is implemented in the QwaveMPS package in a similar manner to the CW pump. We only need to choose the envelope shape with its corresponding parameters, and then it can be added to the Hamiltonian as in the CW case. For example, for a Gaussian $\pi$ pulse (pulse area units, where we expect one perfect Rabi flop in the absence of dephasing): 
\begin{python}

"""Pulse parameters"""
pulse_time=tmax
gaussian_center=1.5 
gaussian_width=0.5 
strength=np.pi

"""Pulse envelope"""
pulsed_pump=strength*qmps.
    gaussian_envelope
    (pulse_time,input_params, 
    gaussian_width,gaussian_center)

"""Choose the Hamiltonian"""
Hm=qmps.hamiltonian_1tls(input_params,
    pulsed_pump)
\end{python}
Again, here we can add detuning by adding an extra parameter at the end of the Hamiltonian (zero by default):
\begin{python}

"""Choose the Hamiltonian"""
Hm=qmps.hamiltonian_1tls(input_params,
    pulsed_pump,delta=np.pi)
\end{python}

Then, the time evolution and observables are calculated in the same way as before. In this example, we have chosen again to start without excitations in the TLS or waveguide, but this is not a method requirement.

In this case, full two-time correlation functions are required [Eqs.~\eqref{eq:G1_twotimes} and~\eqref{eq:G2_twotimes}]. 
These can be calculated using functions with a similar structure to the steady-state case:
\begin{python}
""" First order correlation"""
g1_correls,ts=qmps.correlation_2op_2t
    (corr_bins,a_ops,b_ops,input_params)

""" Second order correlation"""
g2_correls,ts=qmps.correlation_4op_2t
    (corr_bins,a_ops,b_ops,
    c_ops,d_ops,input_params)
\end{python}
and calculate a list of full two-time correlation functions, as well as returning a corresponding list of time points. Again, this function can be used to calculate arbitrary correlation functions of the form $\braket{A(t)B(t+t')C(t+t')D(t)}$, where each of these operators may be a product of bosonic operators to calculate sections of higher-order correlation functions within the two-dimensional time space. Further support also exists for completely general two-time correlation functions of the output field.

In Fig.~\ref{fig:classical_pipulse}, a single TLS in an infinite (a,b) and a semi-infinite waveguide (c,d) is driven with a Gaussian $\pi$ pulse. In the non-Markovian regime, $\gamma \tau = 1$ and $\phi=\pi$. In both cases, the pulse has a width of $\gamma\sigma=0.5$, is centred at $\gamma t_c=1.5$, and has a strength of $\pi$ so that it is a $\pi$ pulse. In 
panels (a) and (c) of Fig.~\ref{fig:classical_pipulse}, population dynamics are calculated, and we can observe the population trapping in the TLS and the feedback loop in the non-Markovian regime, while in the infinite waveguide, the TLS gets populated and then goes back to the ground. 

Panels (b) and (d) of Fig.~\ref{fig:classical_pipulse} show the two-time second-order correlation functions, where we can see the complete two-time dynamics. In Fig.~\ref{fig:classical_pipulse}(b), the correlation of right photons $G^{(2)}_{RR}(t,t+t')$ is shown, where we can see no transmission in the center diagonal, where $t$ is equal to $t+t'$. In Fig.~\ref{fig:classical_pipulse}(d), there is only a single channel with all the photons going out of the loop to the right, so we only have $G^{(2)}(t,t+t')$, with a very different transmission pattern. 

Although not explicitly shown here, time-dependent spectra and spectral intensities can be calculated as well using the following equations~\cite{Liu2024,sofia2025}:
\begin{equation}
\begin{split}
	&\mathcal{S}(\omega,t) = \\
    &\text{Re}\left[ \int_0^{t} dt'' \int_0^{t-t''} \!\! dt'  \langle b^{\dagger}(t'')b(t''+t')\rangle e^{i\Delta_{\omega {\rm c}}t'} \right], 
 \label{eq:Swt}
\end{split}
\end{equation}
and
\begin{equation}
 I(\omega, t) = 
     {\rm Re} \left[ \int_0^{\infty} \! dt'  \braket{b^{\dagger} (t) b(t+t')}  e^{i \Delta _{\rm \omega {\rm c}}t'}    \right]
    \label{eqI},
\end{equation}
where $\Delta_{\omega {\rm c}} = \omega - \omega_{\rm c}$,
with $\omega_{\rm c}$ the central frequency in the rotating frame.
More information about these observables in 
waveguide-QED MPS can be found in~\cite{sofia2025}.

\subsection{Fock-state excitation pulses}
\label{subsec:fock}

\begin{figure*}[ht]
  \centering
  \includegraphics[width=0.9\textwidth]{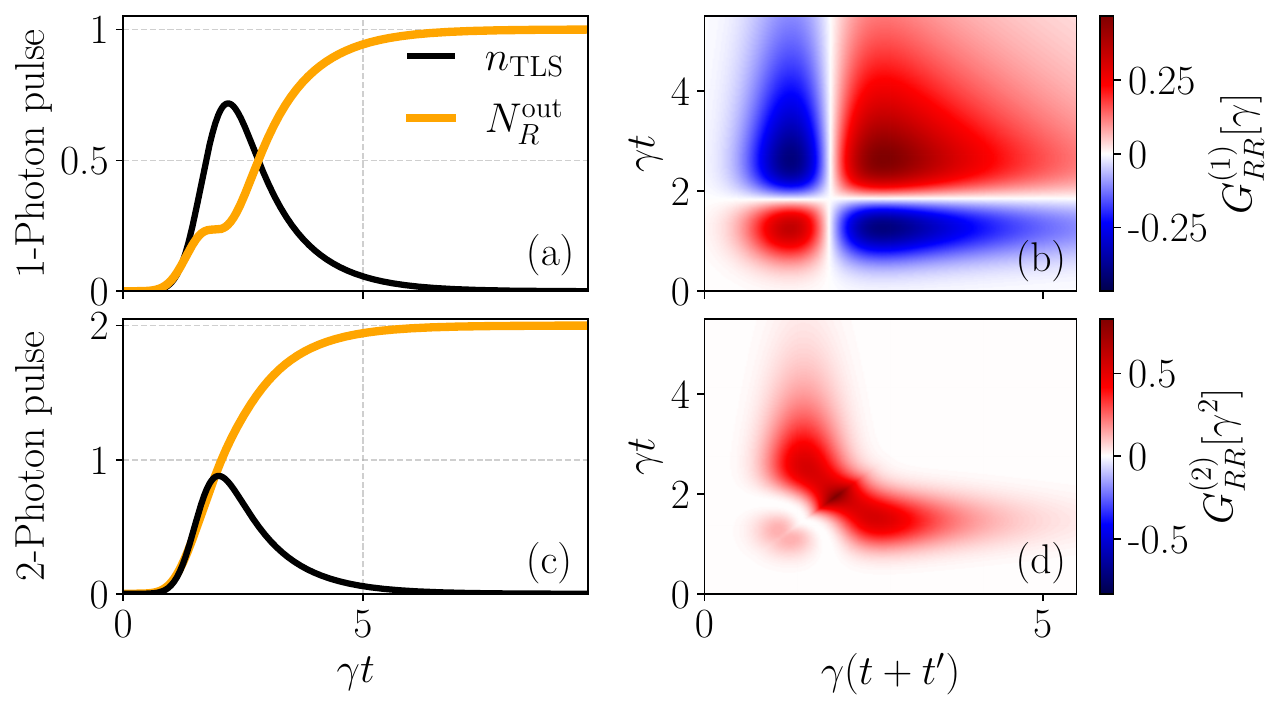}
  \caption{Single chiral TLS driven by a Fock pulse with a Gaussian envelope, with a width of $\gamma\sigma=0.5$ and centred at $\gamma t_c=1.5$. (a,b) Single-photon quantum pulse. (c,d) Two-photon quantum pulse. Population dynamics are shown in (a,c) in black, and the integrated right flux in orange. The first-order correlation function $G^{(1)}_{RR}$ for the 1-photon pulse solution is shown in (b), and the second-order correlation function $G^{(2)}_{RR}$ for the two-photon one is shown in (d).
  }
  \label{fig:quantum_pulse}
\end{figure*}

In our final example, we introduce fully quantum pulses in the Fock state basis, with few-photon pulses now within the waveguide in some chosen direction~\cite{PhysRevA.103.033704}. Thus, we can consider an $n$-photon excitation pulse directly as the initial state, instead of modeling it as a classical pulse in the Hamiltonian. 

Time-bin TN descriptions of photonic states were formulated in Ref.~\cite{Lubasch2018} with the description of product state pulse trains, and time-domain MPS has been used for pulsed nonlinear optics~\cite{Yanagimoto2021} and for propagation through atomic ensembles~\cite{Manzoni2017}. In addition, time-bin and collisional constructions of Fock-state wavepackets~\cite{PhysRevResearch.2.043070,Cilluffo2026}, and multiphoton pulses beyond the two-excitation limit have been treated in Refs.~\cite{PhysRevA.103.033704,PhysRevA.106.023708}; the underlying continuous matrix product description of field states goes back to Ref.~\cite{Verstraete2010}. 
The analytic factorization of an $N$-photon wavepacket into time-bin tensors given below follows automaton-style MPS constructions~\cite{Crosswhite2008}, and is derived for general $N$ in Ref.~\cite{p694-46xh}; MPS simulations of few-photon Fock-state pulses in waveguide-QED were also reported in Ref.~\cite{sofia2025}.

QwaveMPS contributes here to the automation of the Fock state pulse: \texttt{qmps.fock\_pulse} builds the corresponding MPS directly from a photon number, an envelope function, and a propagation direction, and sets the bin dimensions that it requires, so that the tensors below need not be assembled by the user. We give them explicitly for transparency and reproducibility.

Here, we will show the cases of a 1-photon and a 2-photon Fock pulse, but the package can be used to calculate higher numbers of photon pulses as well. 
In the MPS scheme, the 1-photon Fock state is written as,
\begin{equation}
 \ket{\phi_0}_m = \sum_{k=1}^m f_k \Delta B_k^\dagger /\sqrt{\Delta t} \ket{0,...,0}, 
\end{equation}
where $f_k$ corresponds to the pulse envelope at each bin, $k$, and satisfies the normalization condition $\sum_k |f_k|^2 = 1$. This is translated to the following MPS terms ($A_k^{(i_k)}$): 
\begin{itemize}
    \item if $k=1$:
\begin{equation}
    A_1^{(0)}=
    \begin{pmatrix}
        1 & 0
    \end{pmatrix}
    , \ A_1^{(1)}=
    \begin{pmatrix}
        0 & f_1
    \end{pmatrix},
\end{equation}
\item if $1 < k < m$:
\begin{equation}
A_k^{(0)}=
    \begin{pmatrix}
         1 & 0 \\
         0 & 1
    \end{pmatrix}, \ A_k^{(1)}=
    \begin{pmatrix}
        0 &  f_k \\
        0 & 0
    \end{pmatrix},
\end{equation}
\item if $k=m$:
\begin{equation}
A_m^{(0)}=
    \begin{pmatrix}
        0 \\
        1
    \end{pmatrix}, \ A_m^{(1)}=
    \begin{pmatrix}
        f_m \\
        0
    \end{pmatrix},
\end{equation}
\end{itemize}

In the case of having a two-photon initial pulse, 
\begin{equation}
\begin{split}
    &\ket{\phi_0}_m = \frac{1}{\Delta t \sqrt{2}}  
    \Big(\sum_{k=1}^m \big(f_k \Delta B_k^\dagger\big)^2 \\
    &+ 2 \sum_{k=1}^m  \sum_{l=k+1}^m f_k \Delta B_k^\dagger f_l \Delta B_l^\dagger \Big) \ket{0,...,0}, 
\end{split}
\end{equation}  
and the tensors are now:
\begin{itemize}
    \item if $k=1$:
\begin{equation}
\begin{split}
   & A_1^{(0)}=
    \begin{pmatrix}
        1 & 0 & 0
    \end{pmatrix},
    \ A_1^{(1)}=
    \begin{pmatrix}
       0 & f_1 & 0
    \end{pmatrix},
    \\& 
     A_1^{(2)}=
    \begin{pmatrix}
        0 & 0 & \frac{1}{\sqrt{2}} f_1^2
    \end{pmatrix}
    ,
    \end{split}
\end{equation}
\item if $1 < k < m$:
\begin{equation}
\begin{split}
&\hspace{-0.6cm} A_k^{(0)}=
    \begin{pmatrix}
         1 & 0 & 0\\
         0 & 1 & 0 \\
         0 & 0 & 1
    \end{pmatrix}, \ A_k^{(1)}=
    \begin{pmatrix}
        0 & f_k & 0\\
        0 & 0 & f_k \\
        0 & 0 & 0
    \end{pmatrix}, 
    \\
    & \hspace{-0.6cm} A_k^{(2)}=
    \begin{pmatrix}
        0 & 0 & \frac{1}{\sqrt{2}} f_k^2 \\
        0 & 0 & 0 \\
        0 & 0 & 0
    \end{pmatrix},
    \end{split}
\end{equation}
\item if $k=m$:
\begin{equation}
\begin{split}
&A_m^{(0)}=
    \sqrt{2}\begin{pmatrix}
        0 \\
        0 \\
        1
    \end{pmatrix}, \ A_m^{(1)}=
    \sqrt{2}\begin{pmatrix}
        0 \\
        f_m \\
        0
    \end{pmatrix},
    \\
    &\ A_m^{(2)}=
    \sqrt{2}\begin{pmatrix}
        \frac{1}{\sqrt{2}}f_m^2 \\
        0 \\
        0
    \end{pmatrix}.
    \end{split}
\end{equation}
\end{itemize}

Due to the increasing size of the tensors, it is important to carefully choose the proper bin dimensions in each case. For instance, for a single-photon pulse, each channel can have a bin dimension of two, but for the two-photon pulse, we require at least a dimension of three.

This is implemented in a very similar manner as before, but now, instead of choosing the strength of the pulse (such as the Rabi frequency), we need to choose the number of photons that we want in our pulse. For example, for a single-photon Gaussian pulse propagating in the right direction:
\begin{python}
pulse_time = tmax
gaussian_center = 1.5 
gaussian_width = 0.5 
pulse_env=qmps.gaussian_envelope
    (pulse_time,input_params, 
    gaussian_width,gaussian_center)
photon_num = 1

i_n0=qmps.fock_pulse
    (pulse_env,pulse_time,photon_num,
    input_params,direction='R')
\end{python}
The rest of the implementation and calculation of observables follows the same functions as in the previous sections. 

Figure~\ref{fig:quantum_pulse} presents population dynamics and correlation functions of a right chiral TLS in an infinite waveguide interacting with a Gaussian pulse that propagates to the right and contains one (a,b) and two (c,d) photons. In the population dynamics in (a,c), we can see how, in the 1-photon case, the transmitted output flux goes to one after the interaction, since everything is transmitted in the chiral system. However, in (c), the propagating pulse contains two photons, and thus, the integrated flux will go to two at long times. In (b), we present the transmitted first-order correlation function at two times, $G^{(1)}_{RR} (t,t+t')$, where the interaction between the pulse and the TLS is observed. In this case, $G^{(2)}_{RR} (t,t+t')=0$ at all times, since there is only one photon present. In (d), we show the second-order transmitted correlation $G^{(2)}_{RR} (t,t+t')$, which shows the characteristic 
{\em bird-shape} found in the literature~\cite{LeJeannic2022} (in this case, using quantum dot waveguide systems).

Note that more interesting nonlinear correlations can also be explored by varying the time separation between the two temporal wave packets making up the 
two-photon pairs, which need not overlap temporally~\cite{Kozma2026pulsed}.

One of the greatest advantages of using MPS is that we can easily extend this to having a higher number of photons, add additional TLSs, or start with (pre-pulsed) excited TLS, as well as more drives. Such options are rarely possible when using other approaches due to the rapidly increasing Hilbert space or the need for boundary conditions in other discretized methods, like the method of quantum trajectories.  

Table~\ref{table:fock} presents the run times of population dynamics with Fock state pulses and double time correlation functions. As in all the previous cases, population dynamics are calculated very efficiently, with times of around a tenth of a second in both cases. Two-time correlation functions are a bit slower due to the number of points being calculated, but remain efficient, with precise results calculated in under half a minute.

\begin{table}[h]
\centering
\caption{Run times (single computer) for the observables calculated in Sec.~\ref{subsec:fock} with Fock state pulses. Parameters: time step $\gamma \Delta t = 0.05$, final time $\gamma t_{\rm max}=8$, and maximum bond dimension $\chi_{\rm max} = 4$ for the single-photon pulse and $\chi_{\rm max} = 8$ for the 2-photon one. 
}
\begin{tabular}{|c|c|}
    \hline
    Observable  & Run time \\
    \hline
    Populations 1-photon pulse & 0.12 s \\
    Populations 2-photon pulse & 0.19 s \\
    $G^{(1)}(t,t+t')$  & 6.80 s \\
    $G^{(2)}(t,t+t')$  & 28.29 s \\
    \hline
\end{tabular}
\label{table:fock}
\end{table}

\subsection{Scaling and computational requirements}
\label{subsec:scaling}

To characterize the computational demands of a 
genuinely 
{\it non-Markovian, nonlinear
problem}, in this section we use QwaveMPS to benchmark the coherently driven TLS in
front of a mirror introduced in Sec.~\ref{subsubsec:cw_pump}. This is a stringent
test case for two reasons: the excitation number is not conserved, and a
continuously driven emitter feeds photons into the waveguide indefinitely, so no
fixed truncation in the total number of quanta applies; and increasing the delay
lengthens the chain of bins that must be swapped and entangled at every time step,
increasing the MPS entanglement and required maximum bond dimension. Theories
restricted to one or two quanta cannot treat this regime at all; indeed, even for
rather weak fields we find that higher-order correlations are generally important
and informative, such as the four-photon correlations arising with few-photon
squeezing~\cite{sofia2026squeezed}.

\begin{table}[hbt]
\centering
\caption{Scaling benchmark for the coherently driven TLS in a semi-infinite waveguide of Sec.~\ref{subsubsec:cw_pump}, as a function of the drive strength $\Omega$ and the delay time $\tau$, at a fixed mirror phase of $\phi=\pi$. Here $N_{\rm ss}^{\rm loop}$ is the mean photon number in the feedback loop in the steady state, given by Eq.~\eqref{eq:N_in}; $\chi_{\rm max}$ is the converged maximum bond dimension (tested at 8 and multiples of 16). Parameters: time step $\gamma \Delta t = 0.02$ and final time $\gamma t_{\rm max}=20$, run on the single computer workstation described at the start of Sec.~\ref{sec:examples}.}
\label{table:scaling_cw}
\begin{tblr}{
  colspec = {|c|c|c|c|c|c|},
}
\hline
$\Omega/\gamma$ & $\gamma\tau$ & $\chi_{\rm max}$ & $N^{\rm loop}_{\rm ss}$ &
{Run\\time\\{[s]}} & {Peak\\memory\\{[MB]}} \\
\hline
$\pi$  & 1 & 16 & 0.240 & 41.04  & 17.39 \\
$2\pi$ & 1 & 16 & 0.249 & 41.09  & 17.39 \\
$4\pi$ & 1 & 32 & 0.250 & 101.03 & 66.42 \\
$8\pi$ & 1 & 48 & 0.248 & 198.81 & 148.14 \\
\hline
$2\pi$ & 0.5 & 8  & 0.124 & 11.03  & 5.04 \\
$2\pi$ & 1   & 16 & 0.249 & 41.09  & 17.39 \\
$2\pi$ & 2   & 32 & 0.495 & 204.71 & 69.54 \\
$2\pi$ & 4   & 48 & 0.989 & 840.47 & 169.15 \\
\hline
\end{tblr}
\end{table}


Table~\ref{table:scaling_cw} reports, across various pump strengths and delay distances to the mirror, the converged maximum bond dimension, $\chi_{\rm max}$, the steady-state mean number of photons in the feedback loop (calculated by Eq.~\eqref{eq:N_in} at steady-state time) $N_{\rm ss}^{\rm loop}$,  
the simulation run time, and the memory used by the simulation. The values of $\chi_{\rm max}$ quoted here are obtained by increasing the bond-dimension cutoff until the emitter population changes by less than $10^{-3}$ over the full time window, so that they are converged rather than assumed.

The behavior is as expected on physical grounds, but the numbers make the statement quantitative. 
Increasing the drive strength has minimal impact on the mean number of quanta
in the loop; that quantity instead scales linearly with the length of the delay
line, since it reflects the photons emitted over the preceding interval $\tau$.
Increasing the delay lengthens the swapped chain and raises the entanglement across the feedback cut, with $\chi_{\rm max}$ growing from 8 at $\gamma \tau = 0.5$ to 48 at $\gamma \tau = 4$ for the appropriate convergence. In all cases, the peak memory remains below 200~MB; hence, these calculations are still easily accessible on a laptop.

We also mention that the long-time output flux itself shows no simple
dependence on $\Omega$ and $\tau$, with a particularly rich behavior at short delays; a systematic map of this parameter space would be an interesting
application of the package in its own right.

\begin{table}[!ht]
\centering
\caption{Scaling with photon number for $N$-photon Fock-state Gaussian pulses incident on a chiral TLS, as in Sec.~\ref{subsec:fock}. Gaussian envelopes with $\gamma\sigma=0.5$ centered at $\gamma t_c=3$; time step $\gamma \Delta t = 0.05$ and final time $\gamma t_{\rm max}=10$. The memory listed in this table includes only the memory involved in the simulation, and not the Python interpreter/process. 
As with all simulations in this paper, these were run on the single computer workstation described at the start of Sec.~\ref{sec:examples}.}
\label{table:scaling_fock}
\begin{tabular}{|c|c|c|c|}
    \hline
    $N$ $\mkern-8mu$ & $\chi_{\rm max}$ $\mkern-8mu$ & Run time [s] $\mkern-8mu$ & 
     Peak memory [MB] $\mkern-8mu$ \\
    \hline
    1 & 2 & 0.06 & 0.13 \\
    2 & 4 & 0.07 & 0.72 \\
    3 & 6 & 0.10 & 2.06 \\
    4 & 8 & 0.11 & 4.47 \\
    5 & 10 & 0.13 & 8.24 \\
    6 & 12 & 0.18 & 13.67 \\
    \hline
\end{tabular}
\end{table}

Additionally, Table~\ref{table:scaling_fock} gives the converged bond dimension, run time, and simulation memory usage for $N$-photon Fock-state pulses incident on a chiral TLS, as in Sec.~\ref{subsec:fock}, where the photon number is fixed by the input state rather than generated by the drive. In this case, the time bin dimension is fixed by $N+1$ for exact representation of the quantum pulse, and the cost (both time and memory) grows with the photon content of the pulse, both due to the growth of the time bin space and the bond dimension. Population dynamics for pulses containing up to eight photons, obtained with this version of the code, are reported in Ref.~\cite{p694-46xh}.

Finally, we emphasize that the required bond dimension depends on the entanglement generated in a given problem; the tables above are intended as an empirical guide for users choosing \texttt{bond\_max}, and the code raises a warning when the chosen value is too small. 


\subsection{Future Features}
\label{subsec:future}

\phantom{}

In our paper, we have shown selected examples of the capability of using QwaveMPS to solve 
complex waveguide-QED problems, from Markov to non-Markov regimes, from linear to nonlinear regimes, 
including how to compute populations and multi-time quantum correlation functions, all with
run times that are seconds or less.

Yet, there are many other examples and scenarios that could be studied
in future works, with small extensions and future editions of our code. 
In this section, we highlight a few features that we are planning to investigate in the very near future.

Firstly, the problems studied in the previous sections were ideal cases that excluded loss processes (recall MPS does not have any lost photons; everything is treated at the system level). 
This is common for exact methods that solve waveguide-QED problems. 
Hence, a natural extension would be to include off-chip decay and pure dephasing, to get closer to many experimental results~\cite{PhysRevLett.104.017402,PhysRevB.98.045309,PhysRevResearch.3.023168}, since practically every waveguide-QED system has additional sources of loss, including
qubit pure dephasing processes~\cite{PhysRevResearch.3.023168,PhysRevX.10.031011} (for atoms, transmons, flux qubits and quantum dots). We have already implemented this in Ref.~\cite{Matias_2026} using our current version of QwaveMPS, and it will be added as a package-compatible feature in a future version of the package.

Note that although we calculate field correlation functions here, MPS also allows one to calculate atom-atom correlations, and even atom-field correlations, as demonstrated recently in the case of nonlinear superradiance with finite time delays~\cite{slzl-d5dz}. Time-delayed collective effects such as superradiance and subradiance have attracted considerable recent interest~\cite{hmbj-lg7p,PhysRevLett.134.173601,Barahona-Pascual_2026,Capurso2026}, and these correlation features can easily be implemented in future versions of the package. 

Another case that can be easily added is going beyond two coupled TLSs. In fact, we have already studied such cases in other works using MPS, by modeling three coupled emitters, and also well into the deep non-Markovian regime~\cite{PhysRevA.104.L031701,PhysRevA.108.033719,slzl-d5dz}. 

We also stress that our method is not restricted to two-level systems. One can also model three- and four-level quantum dots by increasing the Hilbert subspace of the system part and updating the Hamiltonian accordingly~\cite{PhysRevA.85.043832,Dusanowski2022}. Furthermore, we can study different emitter systems such as Harmonic Oscillators (HOs), transmons (superconducting systems, which behave as anharmonic oscillators)~\cite{PhysRevA.76.042319,blais_circuit_2021,Roth2023}, or systems that go beyond the linear dispersion medium (suitable for studying the ultra-strong regime in waveguide-QED~\cite{SnchezBurillo2015,PhysRevA.106.063717}).
In addition, other initial conditions can be considered. This includes, for example, 
adding initial coherent or squeezed states, or even simulations of $n$-photon bundles and cluster states~\cite{PhysRevA.82.063816,Istrati2020}.

\section{Conclusions}
\label{sec:conclusion}

We have introduced our new software package {\it QwaveMPS}, an efficient open-source Python code for modeling complex waveguide-QED systems, containing waveguide photons and atoms (such as TLSs). The waveguide-QED software package builds on MPS, a numerically exact approach based on tensor-network theories that allows one to discretize the electromagnetic
field into time bins. The waveguide-QED MPS approach allows one to treat waveguide photons exactly and fully account for finite-time delays on the waveguide architecture, and it avoids an exponential growth of the Hilbert space. Not only is it far more powerful than master equation approaches, including non-Markovian ones, but it is also extremely efficient at solving complex many-body interactions for waveguide-QED systems. 
A central contribution of the package is the automation of the waveguide-QED-specific tensor-network operations.

We first gave an overview of MPS for waveguide-QED problems as well as the QwaveMPS package,
and then presented a series of examples, ranging from simple vacuum decay, with and without time-delayed feedback, to complex nonlinear dynamics from multi-quanta excitation, classical field excitation, and excitation by $n$-photon Fock state pulses. 

We then showed how to calculate different observables, starting with time-dependent population dynamics, where we then introduced the entanglement entropy with an example in the nonlinear vacuum dynamics case. 
Steady state solutions of correlations and spectra were studied with a classical CW drive, and the full two-times correlation functions were introduced with pulsed light both in the classical field regime, and later with fully quantum Fock state pulses. Our results were also fully compared and benchmarked with other techniques, where known, showing excellent agreement in all cases. 

Finally, we discussed some ideas for future features, demonstrating how QwaveMPS has the potential to become an important open-source package for the quantum optics community. 

\acknowledgements
This work was supported by the
Natural Sciences and Engineering Research Council of
Canada (NSERC) [Discovery Grants and Quantum Alliance Grants], the Canada Foundation for Innovation
(CFI), and Queen's University, Canada.




\section*{Author contributions}
S.A.R. and M.K. developed the software; S.H. supervised the work; all authors
contributed to the theory and to writing the manuscript.

\section*{AI usage}
 Claude (Anthropic) was used for minor proofreading and language assistance.

\bibliographystyle{quantum}
\bibliography{biblio}

\begin{thebibliography}{100}

\bibitem{PhysRevA.76.062709}
Jung-Tsung Shen and Shanhui Fan.
\newblock ``Strongly correlated multiparticle transport in one dimension through a quantum impurity''.
\newblock \href{https://dx.doi.org/10.1103/PhysRevA.76.062709}{Phys. Rev. A {\bf 76}, 062709}~(2007).

\bibitem{PhysRevLett.98.153003}
Jung-Tsung Shen and Shanhui Fan.
\newblock ``Strongly correlated two-photon transport in a one-dimensional waveguide coupled to a two-level system''.
\newblock \href{https://dx.doi.org/10.1103/PhysRevLett.98.153003}{Phys. Rev. Lett. {\bf 98}, 153003}~(2007).

\bibitem{Witthaut_2010}
D~Witthaut and A~S S\text{\o}rensen.
\newblock ``Photon scattering by a three-level emitter in a one-dimensional waveguide''.
\newblock \href{https://dx.doi.org/10.1088/1367-2630/12/4/043052}{New Journal of Physics {\bf 12}, 043052}~(2010).

\bibitem{PhysRevLett.106.053601}
Dibyendu Roy.
\newblock ``Two-photon scattering by a driven three-level emitter in a one-dimensional waveguide and electromagnetically induced transparency''.
\newblock \href{https://dx.doi.org/10.1103/PhysRevLett.106.053601}{Phys. Rev. Lett. {\bf 106}, 053601}~(2011).

\bibitem{PhysRevLett.113.263604}
E.~Sanchez-Burillo, D.~Zueco, J.~J. Garc\'ia-Ripoll, and L.~Martin-Moreno.
\newblock ``Scattering in the ultrastrong regime: Nonlinear optics with one photon''.
\newblock \href{https://dx.doi.org/10.1103/PhysRevLett.113.263604}{Phys. Rev. Lett. {\bf 113}, 263604}~(2014).

\bibitem{PhysRevLett.116.093601}
Hannes Pichler and Peter Zoller.
\newblock ``Photonic circuits with time delays and quantum feedback''.
\newblock \href{https://dx.doi.org/10.1103/PhysRevLett.116.093601}{Phys. Rev. Lett. {\bf 116}, 093601}~(2016).

\bibitem{Calaj2016}
Giuseppe Calaj\`o, Francesco Ciccarello, Darrick Chang, and Peter Rabl.
\newblock ``Atom-field dressed states in slow-light waveguide {QED}''.
\newblock \href{https://dx.doi.org/10.1103/PhysRevA.93.033833}{Phys. Rev. A {\bf 93}, 033833}~(2016).

\bibitem{Hughes2004}
S.~Hughes.
\newblock ``Enhanced single-photon emission from quantum dots in photonic crystal waveguides and nanocavities''.
\newblock \href{https://dx.doi.org/10.1364/ol.29.002659}{Optics Letters {\bf 29}, 2659}~(2004).

\bibitem{PhysRevLett.120.140404}
Anton~Frisk Kockum, G\"oran Johansson, and Franco Nori.
\newblock ``Decoherence-free interaction between giant atoms in waveguide quantum electrodynamics''.
\newblock \href{https://dx.doi.org/10.1103/PhysRevLett.120.140404}{Phys. Rev. Lett. {\bf 120}, 140404}~(2018).

\bibitem{PhysRevA.87.013820}
Tommaso Tufarelli, Francesco Ciccarello, and M.~S. Kim.
\newblock ``Dynamics of spontaneous emission in a single-end photonic waveguide''.
\newblock \href{https://dx.doi.org/10.1103/PhysRevA.87.013820}{Phys. Rev. A {\bf 87}, 013820}~(2013).

\bibitem{PhysRevLett.118.213601}
A.~Gonz\'alez-Tudela, V.~Paulisch, H.~J. Kimble, and J.~I. Cirac.
\newblock ``Efficient multiphoton generation in waveguide quantum electrodynamics''.
\newblock \href{https://dx.doi.org/10.1103/PhysRevLett.118.213601}{Phys. Rev. Lett. {\bf 118}, 213601}~(2017).

\bibitem{PhysRevA.82.063816}
Huaixiu Zheng, Daniel~J. Gauthier, and Harold~U. Baranger.
\newblock ``Waveguide {QED}: Many-body bound-state effects in coherent and {Fock-state} scattering from a two-level system''.
\newblock \href{https://dx.doi.org/10.1103/PhysRevA.82.063816}{Phys. Rev. A {\bf 82}, 063816}~(2010).

\bibitem{PhysRevA.83.063828}
Paolo Longo, Peter Schmitteckert, and Kurt Busch.
\newblock ``Few-photon transport in low-dimensional systems''.
\newblock \href{https://dx.doi.org/10.1103/PhysRevA.83.063828}{Phys. Rev. A {\bf 83}, 063828}~(2011).

\bibitem{PhysRevA.102.023702}
Juan Rom\'an-Roche, Eduardo S\'anchez-Burillo, and David Zueco.
\newblock ``Bound states in ultrastrong waveguide {QED}''.
\newblock \href{https://dx.doi.org/10.1103/PhysRevA.102.023702}{Phys. Rev. A {\bf 102}, 023702}~(2020).

\bibitem{PhysRevA.100.023834}
Paolo Facchi, Davide Lonigro, Saverio Pascazio, Francesco~V. Pepe, and Domenico Pomarico.
\newblock ``Bound states in the continuum for an array of quantum emitters''.
\newblock \href{https://dx.doi.org/10.1103/PhysRevA.100.023834}{Phys. Rev. A {\bf 100}, 023834}~(2019).

\bibitem{Dinc2019}
Fatih Dinc and Agata~M. Bra{\'{n}}czyk.
\newblock ``Non-{Markovian} super-superradiance in a linear chain of up to 100 qubits''.
\newblock \href{https://dx.doi.org/10.1103/physrevresearch.1.032042}{Phys. Rev. Research {\bf 1}, 032042}~(2019).

\bibitem{PhysRevLett.122.073601}
Giuseppe Calaj\`o, Yao-Lung~L. Fang, Harold~U. Baranger, and Francesco Ciccarello.
\newblock ``Exciting a bound state in the continuum through multiphoton scattering plus delayed quantum feedback''.
\newblock \href{https://dx.doi.org/10.1103/PhysRevLett.122.073601}{Phys. Rev. Lett. {\bf 122}, 073601}~(2019).

\bibitem{GonzalezBallestero2013}
C~Gonzalez-Ballestero, F~J Garc\'ia-Vidal, and Esteban Moreno.
\newblock ``Non-{Markovian} effects in waveguide-mediated entanglement''.
\newblock \href{https://dx.doi.org/10.1088/1367-2630/15/7/073015}{New Journal of Physics {\bf 15}, 073015}~(2013).

\bibitem{RevModPhys.89.021001}
Dibyendu Roy, C.~M. Wilson, and Ofer Firstenberg.
\newblock ``Colloquium: Strongly interacting photons in one-dimensional continuum''.
\newblock \href{https://dx.doi.org/10.1103/RevModPhys.89.021001}{Rev. Mod. Phys. {\bf 89}, 021001}~(2017).

\bibitem{Kannan2020}
Bharath Kannan, Max~J. Ruckriegel, Daniel~L. Campbell, Anton Frisk~Kockum, Jochen Braum\"{u}ller, David~K. Kim, Morten Kjaergaard, Philip Krantz, Alexander Melville, Bethany~M. Niedzielski, Antti Veps\"{a}l\"{a}inen, Roni Winik, Jonilyn~L. Yoder, Franco Nori, Terry~P. Orlando, Simon Gustavsson, and William~D. Oliver.
\newblock ``Waveguide quantum electrodynamics with superconducting artificial giant atoms''.
\newblock \href{https://dx.doi.org/10.1038/s41586-020-2529-9}{Nature {\bf 583}, 775--779}~(2020).

\bibitem{doi:10.1126/sciadv.aaw0297}
M.~Bello, G.~Platero, J.~I. Cirac, and A.~Gonz\'alez-Tudela.
\newblock ``Unconventional quantum optics in topological waveguide {QED}''.
\newblock \href{https://dx.doi.org/10.1126/sciadv.aaw0297}{Science Advances {\bf 5}, eaaw0297}~(2019).

\bibitem{PhysRevResearch.6.L032017}
Maria Maffei, Domenico Pomarico, Paolo Facchi, Giuseppe Magnifico, Saverio Pascazio, and Francesco~V. Pepe.
\newblock ``Directional emission and photon bunching from a qubit pair in waveguide''.
\newblock \href{https://dx.doi.org/10.1103/PhysRevResearch.6.L032017}{Phys. Rev. Res. {\bf 6}, L032017}~(2024).

\bibitem{Lodahl2017}
Peter Lodahl, Sahand Mahmoodian, S{\o}ren Stobbe, Arno Rauschenbeutel, Philipp Schneeweiss, J\"{u}rgen Volz, Hannes Pichler, and Peter Zoller.
\newblock ``Chiral quantum optics''.
\newblock \href{https://dx.doi.org/10.1038/nature21037}{Nature {\bf 541}, 473--480}~(2017).

\bibitem{RevModPhys.95.015002}
Alexandra~S. Sheremet, Mihail~I. Petrov, Ivan~V. Iorsh, Alexander~V. Poshakinskiy, and Alexander~N. Poddubny.
\newblock ``Waveguide quantum electrodynamics: Collective radiance and photon-photon correlations''.
\newblock \href{https://dx.doi.org/10.1103/RevModPhys.95.015002}{Rev. Mod. Phys. {\bf 95}, 015002}~(2023).

\bibitem{BROWNE20172}
Dan Browne, Sougato Bose, Florian Mintert, and M.S. Kim.
\newblock ``From quantum optics to quantum technologies''.
\newblock \href{https://dx.doi.org/https://doi.org/10.1016/j.pquantelec.2017.06.002}{Progress in Quantum Electronics {\bf 54}, 2--18}~(2017).

\bibitem{Wang2025}
Hui Wang, Timothy~C. Ralph, Jelmer~J. Renema, Chao-Yang Lu, and Jian-Wei Pan.
\newblock ``Scalable photonic quantum technologies''.
\newblock \href{https://dx.doi.org/10.1038/s41563-025-02306-7}{Nature Materials {\bf 24}, 1883--1897}~(2025).

\bibitem{gardiner_zoller_2010}
Crispin~W Gardiner and Peter Zoller.
\newblock ``Quantum noise: a handbook of {Markovian} and non-{Markovian} quantum stochastic methods with applications to quantum optics''.
\newblock Springer, Berlin~(2010).

\bibitem{PhysRevResearch.3.023030}
Sofia Arranz~Regidor, Gavin Crowder, Howard Carmichael, and Stephen Hughes.
\newblock ``Modeling quantum light-matter interactions in waveguide {QED} with retardation, nonlinear interactions, and a time-delayed feedback: Matrix product states versus a space-discretized waveguide model''.
\newblock \href{https://dx.doi.org/10.1103/PhysRevResearch.3.023030}{Phys. Rev. Res. {\bf 3}, 023030}~(2021).

\bibitem{PhysRevA.106.023708}
Kisa Barkemeyer, Andreas Knorr, and Alexander Carmele.
\newblock ``Heisenberg treatment of multiphoton pulses in waveguide {QED} with time-delayed feedback''.
\newblock \href{https://dx.doi.org/10.1103/PhysRevA.106.023708}{Phys. Rev. A {\bf 106}, 023708}~(2022).

\bibitem{PhysRevA.82.063821}
Shanhui Fan, S.~E. Kocaba\mbox{\c{s}}, and Jung-Tsung Shen.
\newblock ``Input-output formalism for few-photon transport in one-dimensional nanophotonic waveguides coupled to a qubit''.
\newblock \href{https://dx.doi.org/10.1103/PhysRevA.82.063821}{Phys. Rev. A {\bf 82}, 063821}~(2010).

\bibitem{Rephaeli2012FewPhotonSC}
Eden Rephaeli and Shanhui Fan.
\newblock ``Few-photon single-atom cavity {QED} with input-output formalism in {Fock} space''.
\newblock IEEE Journal of Selected Topics in Quantum Electronics {\bf 18}, 1754--1762~(2012).
\newblock  url:~\url{https://api.semanticscholar.org/CorpusID:40913017}.

\bibitem{PhysRevA.101.023807}
Gavin Crowder, J.~Howard Carmichael, and Stephen Hughes.
\newblock ``Quantum trajectory theory of few-photon {cavity-{QED}} systems with a time-delayed coherent feedback''.
\newblock \href{https://dx.doi.org/10.1103/PhysRevA.101.023807}{Phys. Rev. A {\bf 101}, 023807}~(2020).

\bibitem{Nysteen2015}
Anders Nysteen, Philip~Tr\text{\o}st Kristensen, Dara P~S McCutcheon, Per Kaer, and Jesper M\text{\o}rk.
\newblock ``Scattering of two photons on a quantum emitter in a one-dimensional waveguide: exact dynamics and induced correlations''.
\newblock \href{https://dx.doi.org/10.1088/1367-2630/17/2/023030}{New Journal of Physics {\bf 17}, 023030}~(2015).

\bibitem{Chen_2011}
Yuntian Chen, Martijn Wubs, Jesper M\text{\o}rk, and A~Femius Koenderink.
\newblock ``Coherent single-photon absorption by single emitters coupled to one-dimensional nanophotonic waveguides''.
\newblock \href{https://dx.doi.org/10.1088/1367-2630/13/10/103010}{New Journal of Physics {\bf 13}, 103010}~(2011).

\bibitem{chen2026resonancefluorescenceartificialatom}
Ching-Yeh Chen, Gavin Crowder, Zheng-Qi Niu, Ping~Yi Wen, Yen-Hsiang Lin, Jeng-Chung Chen, Zhi-Rong Lin, Franco Nori, Stephen Hughes, and Io-Chun Hoi.
\newblock ``Resonance fluorescence of an artificial atom with a time-delayed coherent feedback''~(2026).
\newblock  \href{http://arxiv.org/abs/2603.28004}{arXiv:2603.28004}.

\bibitem{You2011}
J.~Q. You and Franco Nori.
\newblock ``Atomic physics and quantum optics using superconducting circuits''.
\newblock \href{https://dx.doi.org/10.1038/nature10122}{Nature {\bf 474}, 589--597}~(2011).

\bibitem{blais_circuit_2021}
Alexandre Blais, Arne~L. Grimsmo, S.~M. Girvin, and Andreas Wallraff.
\newblock ``Circuit quantum electrodynamics''.
\newblock \href{https://dx.doi.org/10.1103/revmodphys.93.025005}{Reviews of Modern Physics~{\bf 93}}~(2021).

\bibitem{PhysRevX.14.031055}
Srujan Meesala, David Lake, Steven Wood, Piero Chiappina, Changchun Zhong, Andrew~D. Beyer, Matthew~D. Shaw, Liang Jiang, and Oskar Painter.
\newblock ``Quantum entanglement between optical and microwave photonic qubits''.
\newblock \href{https://dx.doi.org/10.1103/PhysRevX.14.031055}{Phys. Rev. X {\bf 14}, 031055}~(2024).

\bibitem{Marcaud2025}
Guillaume Marcaud, David Perello, Cliff Chen, Esha Umbarkar, Conan Weiland, Jiansong Gao, Sandra Diez, Victor Ly, Neha Mahuli, Nathan D'Souza, Yuan He, Shahriar Aghaeimeibodi, Rachel Resnick, Cherno Jaye, Abdul~K. Rumaiz, Daniel~A. Fischer, Matthew Hunt, Oskar Painter, and Ignace Jarrige.
\newblock ``Low-loss superconducting resonators fabricated from tantalum films grown at room temperature''.
\newblock \href{https://dx.doi.org/10.1038/s43246-025-00897-x}{Communications Materials{\bf \ 6}}~(2025).

\bibitem{7rnm-rxhh}
Aniket Maiti, John~W.O. Garmon, Yao Lu, Alessandro Miano, Luigi Frunzio, and Robert~J. Schoelkopf.
\newblock ``Linear quantum coupler for clean bosonic control''.
\newblock \href{https://dx.doi.org/10.1103/7rnm-rxhh}{PRX Quantum {\bf 6}, 040326}~(2025).

\bibitem{PhysRevResearch.5.033155}
N.~Janzen, X.~Dai, S.~Ren, J.~Shi, and A.~Lupascu.
\newblock ``Tunable coupler for mediating interactions between a two-level system and a waveguide from a decoupled state to the ultrastrong coupling regime''.
\newblock \href{https://dx.doi.org/10.1103/PhysRevResearch.5.033155}{Phys. Rev. Res. {\bf 5}, 033155}~(2023).

\bibitem{LeJeannic2022}
Hanna Le~Jeannic, Alexey Tiranov, Jacques Carolan, Tom\'as Ramos, Ying Wang, Martin~Hayhurst Appel, Sven Scholz, Andreas~D. Wieck, Arne Ludwig, Nir Rotenberg, Leonardo Midolo, Juan~Jos\'e Garc\'ia-Ripoll, Anders~S. S{\o}rensen, and Peter Lodahl.
\newblock ``Dynamical photon-photon interaction mediated by a quantum emitter''.
\newblock \href{https://dx.doi.org/10.1038/s41567-022-01720-x}{Nature Physics {\bf 18}, 1191--1195}~(2022).

\bibitem{PhysRevLett.126.023603}
Hanna Le~Jeannic, Tom\'as Ramos, Signe~F. Simonsen, Tommaso Pregnolato, Zhe Liu, R\"udiger Schott, Andreas~D. Wieck, Arne Ludwig, Nir Rotenberg, Juan~Jos\'e Garc\'{\i}a-Ripoll, and Peter Lodahl.
\newblock ``Experimental reconstruction of the few-photon nonlinear scattering matrix from a single quantum dot in a nanophotonic waveguide''.
\newblock \href{https://dx.doi.org/10.1103/PhysRevLett.126.023603}{Phys. Rev. Lett. {\bf 126}, 023603}~(2021).

\bibitem{Patel2026single}
Tarun Patel, Matteo Pennacchietti, Sayan Gangopadhyay, Dan Dalacu, Philip Poole, Stephen Hughes, and Michael Reimer.
\newblock ``Single photon reflection from a quantum dot in a waveguide''.
\newblock Nature Communications, in press~(2026).
\newblock \href{https://doi.org/10.21203/rs.3.rs-8979997/v1}{preprint}.

\bibitem{PhysRevA.106.063717}
Sergi Terradas-Brians\'o, Carlos~A. Gonz\'alez-Guti\'errez, Franco Nori, Luis Mart\'{\i}n-Moreno, and David Zueco.
\newblock ``Ultrastrong waveguide {QED} with giant atoms''.
\newblock \href{https://dx.doi.org/10.1103/PhysRevA.106.063717}{Phys. Rev. A {\bf 106}, 063717}~(2022).

\bibitem{PhysRevResearch.2.043184}
Angelo Carollo, Dario Cilluffo, and Francesco Ciccarello.
\newblock ``Mechanism of decoherence-free coupling between giant atoms''.
\newblock \href{https://dx.doi.org/10.1103/PhysRevResearch.2.043184}{Phys. Rev. Research {\bf 2}, 043184}~(2020).

\bibitem{PhysRevResearch.2.043070}
Dario Cilluffo, Angelo Carollo, Salvatore Lorenzo, Jonathan~A. Gross, G.~Massimo Palma, and Francesco Ciccarello.
\newblock ``Collisional picture of quantum optics with giant emitters''.
\newblock \href{https://dx.doi.org/10.1103/PhysRevResearch.2.043070}{Phys. Rev. Res. {\bf 2}, 043070}~(2020).

\bibitem{Wang_2022}
Xin Wang and Hong-Rong Li.
\newblock ``Chiral quantum network with giant atoms''.
\newblock \href{https://dx.doi.org/10.1088/2058-9565/ac6a04}{Quantum Science and Technology {\bf 7}, 035007}~(2022).

\bibitem{Liu:22}
Ning Liu, Xin Wang, Xia Wang, Xiao-San Ma, and Mu-Tian Cheng.
\newblock ``Tunable single photon nonreciprocal scattering based on giant atom-waveguide chiral couplings''.
\newblock \href{https://dx.doi.org/10.1364/OE.460255}{Opt. Express {\bf 30}, 23428--23438}~(2022).

\bibitem{PhysRevResearch.6.013279}
Xin Wang, Huai-Bing Zhu, Tao Liu, and Franco Nori.
\newblock ``Realizing quantum optics in structured environments with giant atoms''.
\newblock \href{https://dx.doi.org/10.1103/PhysRevResearch.6.013279}{Phys. Rev. Res. {\bf 6}, 013279}~(2024).

\bibitem{PhysRevA.93.062104}
Tom\'as Ramos, Beno\^{\i}t Vermersch, Philipp Hauke, Hannes Pichler, and Peter Zoller.
\newblock ``Non-markovian dynamics in chiral quantum networks with spins and photons''.
\newblock \href{https://dx.doi.org/10.1103/PhysRevA.93.062104}{Phys. Rev. A {\bf 93}, 062104}~(2016).

\bibitem{hmbj-lg7p}
Irene Papaefstathiou, Daniel Malz, J.~Ignacio Cirac, and Mari~Carmen Ba\~nuls.
\newblock ``Efficient tensor-network simulation of multiemitter non-{M}arkovian systems''.
\newblock \href{https://dx.doi.org/10.1103/hmbj-lg7p}{Phys. Rev. A {\bf 112}, 013721}~(2025).

\bibitem{PhysRevX.14.031043}
Kseniia Vodenkova and Hannes Pichler.
\newblock ``Continuous coherent quantum feedback with time delays: Tensor network solution''.
\newblock \href{https://dx.doi.org/10.1103/PhysRevX.14.031043}{Phys. Rev. X {\bf 14}, 031043}~(2024).

\bibitem{Capurso2026}
Rosa~Lucia Capurso, Giuseppe Calaj\`o, Simone Montangero, Saverio Pascazio, Francesco~V. Pepe, Maria Maffei, Giuseppe Magnifico, and Paolo Facchi.
\newblock ``Superradiant decay in non-{M}arkovian waveguide quantum electrodynamics''.
\newblock \href{https://dx.doi.org/10.1038/s42005-026-02722-4}{Communications Physics}~(2026).

\bibitem{mcculloch_density-matrix_2007}
Ian~P McCulloch.
\newblock ``From density-matrix renormalization group to matrix product states''.
\newblock \href{https://dx.doi.org/10.1088/1742-5468/2007/10/p10014}{Journal of Statistical Mechanics: Theory and Experiment {\bf 2007}, P10014--P10014}~(2007).

\bibitem{Jaderberg:2025iei}
Ben Jaderberg, George Pennington, Kate~V. Marshall, Lewis~W. Anderson, Abhishek Agarwal, Lachlan~P. Lindoy, Ivan Rungger, Stefano Mensa, and Jason Crain.
\newblock ``Variational preparation of normal matrix product states on quantum computers''.
\newblock \href{https://dx.doi.org/10.1103/m5kb-4f43}{Phys. Rev. Res. {\bf 8}, 013081}~(2026).

\bibitem{PhysRevA.109.062437}
Baptiste Anselme~Martin, Thomas Ayral, Fran\ifmmode\mbox{\c{c}}\else\c{c}\fi{}ois Jamet, Marko~J. Ran\ifmmode \check{c}\else \v{c}\fi{}i\ifmmode~\acute{c}\else \'{c}\fi{}, and Pascal Simon.
\newblock ``Combining matrix product states and noisy quantum computers for quantum simulation''.
\newblock \href{https://dx.doi.org/10.1103/PhysRevA.109.062437}{Phys. Rev. A {\bf 109}, 062437}~(2024).

\bibitem{Manzoni2017}
Marco~T. Manzoni, Darrick~E. Chang, and James~S. Douglas.
\newblock ``Simulating quantum light propagation through atomic ensembles using matrix product states''.
\newblock \href{https://dx.doi.org/10.1038/s41467-017-01416-4}{Nature Communications {\bf 8}, 1743}~(2017).

\bibitem{johansson2012qutip}
J.R. Johansson, P.D. Nation, and Franco Nori.
\newblock ``{QuTiP}: An open-source python framework for the dynamics of open quantum systems''.
\newblock \href{https://dx.doi.org/https://doi.org/10.1016/j.cpc.2012.02.021}{Computer Physics Communications {\bf 183}, 1760--1772}~(2012).

\bibitem{johansson_qutip_2013}
J.~R. Johansson, P.~D. Nation, and Franco Nori.
\newblock ``{QuTiP} 2: {A} {Python} framework for the dynamics of open quantum systems''.
\newblock \href{https://dx.doi.org/10.1016/j.cpc.2012.11.019}{Computer Physics Communications {\bf 184}, 1234--1240}~(2013).

\bibitem{BundgaardNielsen2025waveguideqedjl}
Matias Bundgaard-Nielsen, Dirk Englund, Mikkel Heuck, and Stefan Krastanov.
\newblock ``Waveguide{QED}.jl: {A}n {E}fficient {F}ramework for {S}imulating {N}on-{M}arkovian {W}aveguide {Q}uantum {E}lectrodynamics''.
\newblock \href{https://dx.doi.org/10.22331/q-2025-04-17-1710}{{Quantum} {\bf 9}, 1710}~(2025).

\bibitem{Dinc2019exactmarkoviannon}
Fatih Dinc, {\.{I}}lke Ercan, and Agata~M. Bra{\'{n}}czyk.
\newblock ``Exact {M}arkovian and non-{M}arkovian time dynamics in waveguide {QED}: collective interactions, bound states in continuum, superradiance and subradiance''.
\newblock \href{https://dx.doi.org/10.22331/q-2019-12-09-213}{{Quantum} {\bf 3}, 213}~(2019).

\bibitem{PhysRevA.82.033804}
Eden Rephaeli, Jung-Tsung Shen, and Shanhui Fan.
\newblock ``Full inversion of a two-level atom with a single-photon pulse in one-dimensional geometries''.
\newblock \href{https://dx.doi.org/10.1103/PhysRevA.82.033804}{Phys. Rev. A {\bf 82}, 033804}~(2010).

\bibitem{Pichler11362}
Hannes Pichler, Soonwon Choi, Peter Zoller, and Mikhail~D. Lukin.
\newblock ``Universal photonic quantum computation via time-delayed feedback''.
\newblock \href{https://dx.doi.org/10.1073/pnas.1711003114}{Proceedings of the National Academy of Sciences {\bf 114}, 11362--11367}~(2017).

\bibitem{orus_practical_2014}
Rom{\'a}n Or{\'u}s.
\newblock ``A practical introduction to tensor networks: Matrix product states and projected entangled pair states''.
\newblock \href{https://dx.doi.org/10.1016/j.aop.2014.06.013}{Annals of Physics {\bf 349}, 117--158}~(2014).

\bibitem{githubQwaveMPS}
``{G}it{H}ub - {Q}wave{M}{P}{S}: {P}ython package to solve waveguide {Q}{E}{D} systems using {M}atrix {P}roduct {S}tates --- github.com''.
\newblock \url{https://github.com/QwaveMPS/QwaveMPS}~(2026).

\bibitem{itensor}
Matthew Fishman, Steven~R. White, and E.~Miles Stoudenmire.
\newblock ``{The ITensor Software Library for Tensor Network Calculations}''.
\newblock \href{https://dx.doi.org/10.21468/SciPostPhysCodeb.4}{SciPost Phys. CodebasesPage~4}~(2022).

\bibitem{Suess2017}
Daniel Suess and Milan Holz\"apfel.
\newblock ``{mpnum}: A matrix product representation library for {Python}''.
\newblock \href{https://dx.doi.org/10.21105/joss.00465}{Journal of Open Source Software {\bf 2}, 465}~(2017).

\bibitem{mpsdynamicsjl_2024}
Thibaut Lacroix, Brieuc Le~D\'e, Angela Riva, Angus~J. Dunnett, and Alex~W. Chin.
\newblock ``{MPSDynamics}.jl: {Tensor} network simulations for finite-temperature (non-{Markovian}) open quantum system dynamics''.
\newblock \href{https://dx.doi.org/10.1063/5.0223107}{The Journal of Chemical Physics {\bf 161}, 084116}~(2024).

\bibitem{Liu2025}
Yuan Liu, Hong-Bo Sun, and Linhan Lin.
\newblock ``Suppression of local decay rate through energy quantum confinement effect in non-{Markovian} waveguide {QED}''.
\newblock \href{https://dx.doi.org/10.1186/s43074-025-00167-6}{PhotoniX {\bf 6}, 7}~(2025).

\bibitem{Guimond_2017}
P-O Guimond, M~Pletyukhov, H~Pichler, and P~Zoller.
\newblock ``Delayed coherent quantum feedback from a scattering theory and a matrix product state perspective''.
\newblock \href{https://dx.doi.org/10.1088/2058-9565/aa7f03}{Quantum Science and Technology {\bf 2}, 044012}~(2017).

\bibitem{slzl-d5dz}
Sofia Arranz~Regidor, Franco Nori, and Stephen Hughes.
\newblock ``Theory of multiqubit superradiance in a waveguide in the presence of finite delay times and two quantum excitations''.
\newblock \href{https://dx.doi.org/10.1103/slzl-d5dz}{Phys. Rev. A {\bf 112}, 063702}~(2025).

\bibitem{Grimsmo2015}
Arne~L. Grimsmo.
\newblock ``Time-delayed quantum feedback control''.
\newblock \href{https://dx.doi.org/10.1103/PhysRevLett.115.060402}{Phys. Rev. Lett. {\bf 115}, 060402}~(2015).

\bibitem{PhysRevA.106.013714}
Gavin Crowder, Lora Ramunno, and Stephen Hughes.
\newblock ``Quantum trajectory theory and simulations of nonlinear spectra and multiphoton effects in {waveguide-{QED}} systems with a time-delayed coherent feedback''.
\newblock \href{https://dx.doi.org/10.1103/PhysRevA.106.013714}{Phys. Rev. A {\bf 106}, 013714}~(2022).

\bibitem{Liu2024}
Shunfa Liu, Chris Gustin, Hanqing Liu, Xueshi Li, Ying Yu, Haiqiao Ni, Zhichuan Niu, Stephen Hughes, Xuehua Wang, and Jin Liu.
\newblock ``Dynamic resonance fluorescence in solid-state cavity quantum electrodynamics''.
\newblock \href{https://dx.doi.org/10.1038/s41566-023-01359-x}{Nature Photonics}~(2024).

\bibitem{sofia2025}
Sofia Arranz~Regidor, Andreas Knorr, and Stephen Hughes.
\newblock ``Theory and simulations of few-photon {Fock} state pulses strongly interacting with a single qubit in a waveguide: {Exact} population dynamics and time-dependent spectra''.
\newblock \href{https://dx.doi.org/10.1103/lp1b-yswm}{Physical Review Research {\bf 7}, 23295}~(2025).

\bibitem{PhysRevA.103.033704}
Kisa Barkemeyer, Andreas Knorr, and Alexander Carmele.
\newblock ``Strongly entangled system-reservoir dynamics with multiphoton pulses beyond the two-excitation limit: Exciting the atom-photon bound state''.
\newblock \href{https://dx.doi.org/10.1103/PhysRevA.103.033704}{Phys. Rev. A {\bf 103}, 033704}~(2021).

\bibitem{Lubasch2018}
Michael Lubasch, Antonio~A. Valido, Jelmer~J. Renema, W.~Steven Kolthammer, Dieter Jaksch, M.~S. Kim, Ian Walmsley, and Ra\'{u}l Garc\'{\i}a-Patr\'{o}n.
\newblock ``Tensor network states in time-bin quantum optics''.
\newblock \href{https://dx.doi.org/10.1103/PhysRevA.97.062304}{Phys. Rev. A {\bf 97}, 062304}~(2018).

\bibitem{Yanagimoto2021}
Ryotatsu Yanagimoto, Edwin Ng, Logan~G. Wright, Tatsuhiro Onodera, and Hideo Mabuchi.
\newblock ``Efficient simulation of ultrafast quantum nonlinear optics with matrix product states''.
\newblock \href{https://dx.doi.org/10.1364/OPTICA.423044}{Optica {\bf 8}, 1306--1315}~(2021).

\bibitem{Cilluffo2026}
Dario Cilluffo, Luca Ferialdi, G.~Massimo Palma, Giuseppe Calaj\`o, and Francesco Ciccarello.
\newblock ``Multimode-cavity picture of non-{M}arkovian waveguide {QED}''.
\newblock \href{https://dx.doi.org/10.1103/PhysRevResearch.8.023172}{Phys. Rev. Research {\bf 8}, 023172}~(2026).

\bibitem{Verstraete2010}
F.~Verstraete and J.~I. Cirac.
\newblock ``Continuous matrix product states for quantum fields''.
\newblock \href{https://dx.doi.org/10.1103/PhysRevLett.104.190405}{Phys. Rev. Lett. {\bf 104}, 190405}~(2010).

\bibitem{Crosswhite2008}
Gregory~M. Crosswhite and Dave Bacon.
\newblock ``Finite automata for caching in matrix product algorithms''.
\newblock \href{https://dx.doi.org/10.1103/PhysRevA.78.012356}{Phys. Rev. A {\bf 78}, 012356}~(2008).

\bibitem{p694-46xh}
Sofia Arranz~Regidor, Matthew Kozma, and Stephen Hughes.
\newblock ``Quantum dynamics of few-photon pulsed waveguide {QED} with a single artificial atom: Frequency-dependent scattering theory and time-dependent matrix product states''.
\newblock \href{https://dx.doi.org/10.1103/p694-46xh}{Phys. Rev. Res. {\bf 8}, 033260}~(2026).

\bibitem{Kozma2026pulsed}
Matthew Kozma, Sofia {Arranz Regidor}, and Stephen Hughes.
\newblock ``Pulsed two-photon scattering from a single atom in a waveguide with delay-modified temporal correlations''.
\newblock \href{https://dx.doi.org/10.1364/OPTICAQ.XXXXXX}{Optica Quantum {\bf 4}, 435}~(2026).

\bibitem{sofia2026squeezed}
Sofia Arranz~Regidor, Matthew Kozma, and Stephen Hughes.
\newblock ``Matrix product state theory of few-photon squeezed pulses interacting with a two-level emitter in a waveguide''~(2026).
\newblock  \href{http://arxiv.org/abs/2608.06590}{arXiv:2608.06590}.

\bibitem{PhysRevLett.104.017402}
A.~J. Ramsay, Achanta~Venu Gopal, E.~M. Gauger, A.~Nazir, B.~W. Lovett, A.~M. Fox, and M.~S. Skolnick.
\newblock ``Damping of exciton {Rabi} rotations by acoustic phonons in optically excited $\mathrm{InGaAs}/\mathrm{GaAs}$ quantum dots''.
\newblock \href{https://dx.doi.org/10.1103/PhysRevLett.104.017402}{Phys. Rev. Lett. {\bf 104}, 017402}~(2010).

\bibitem{PhysRevB.98.045309}
Chris Gustin and Stephen Hughes.
\newblock ``Pulsed excitation dynamics in quantum-dot--cavity systems: Limits to optimizing the fidelity of on-demand single-photon sources''.
\newblock \href{https://dx.doi.org/10.1103/PhysRevB.98.045309}{Phys. Rev. B {\bf 98}, 045309}~(2018).

\bibitem{PhysRevResearch.3.023168}
Oliver Kaestle, Regina Finsterhoelzl, Andreas Knorr, and Alexander Carmele.
\newblock ``{Continuous and time-discrete non-Markovian system-reservoir interactions: Dissipative coherent quantum feedback in Liouville space}''.
\newblock \href{https://dx.doi.org/10.1103/PhysRevResearch.3.023168}{Phys. Rev. Res. {\bf 3}, 023168}~(2021).

\bibitem{PhysRevX.10.031011}
Sahand Mahmoodian, Giuseppe Calaj\`o, Darrick~E. Chang, Klemens Hammerer, and Anders~S. S\o{}rensen.
\newblock ``Dynamics of many-body photon bound states in chiral waveguide {QED}''.
\newblock \href{https://dx.doi.org/10.1103/PhysRevX.10.031011}{Phys. Rev. X {\bf 10}, 031011}~(2020).

\bibitem{Matias_2026}
Matias Bundgaard-Nielsen, Matthew Kozma, Sofia~Arranz Regidor, and Stephen Hughes.
\newblock ``Decoherence in waveguide quantum electrodynamics using matrix product states''.
\newblock \href{https://dx.doi.org/10.1103/fmpt-dp44}{Physical Review Research, in-press}~(2026).

\bibitem{PhysRevLett.134.173601}
Bennet Windt, Miguel Bello, Daniel Malz, and J.~Ignacio Cirac.
\newblock ``Effects of retardation on many-body superradiance in chiral waveguide {QED}''.
\newblock \href{https://dx.doi.org/10.1103/PhysRevLett.134.173601}{Phys. Rev. Lett. {\bf 134}, 173601}~(2025).

\bibitem{Barahona-Pascual_2026}
Carlos Barahona-Pascual, Hong Jiang, Alan~C Santos, and Juan~Jos\'e Garc\'ia-Ripoll.
\newblock ``Time-delayed collective dynamics in waveguide {QED} and bosonic quantum networks''.
\newblock \href{https://dx.doi.org/10.1088/2058-9565/ae2291}{Quantum Science and Technology {\bf 11}, 015016}~(2025).

\bibitem{PhysRevA.104.L031701}
Sofia Arranz~Regidor and Stephen Hughes.
\newblock ``Cavitylike strong coupling in macroscopic waveguide {QED} using three coupled qubits in the deep non-{Markovian} regime''.
\newblock \href{https://dx.doi.org/10.1103/PhysRevA.104.L031701}{Phys. Rev. A {\bf 104}, L031701}~(2021).

\bibitem{PhysRevA.108.033719}
Sofia Arranz~Regidor and Stephen Hughes.
\newblock ``Probing dressed states and quantum nonlinearities in a strongly coupled three-qubit waveguide system under optical pumping''.
\newblock \href{https://dx.doi.org/10.1103/PhysRevA.108.033719}{Phys. Rev. A {\bf 108}, 033719}~(2023).

\bibitem{PhysRevA.85.043832}
Huaixiu Zheng, Daniel~J. Gauthier, and Harold~U. Baranger.
\newblock ``Strongly correlated photons generated by coupling a three- or four-level system to a waveguide''.
\newblock \href{https://dx.doi.org/10.1103/PhysRevA.85.043832}{Phys. Rev. A {\bf 85}, 043832}~(2012).

\bibitem{Dusanowski2022}
Lukasz Dusanowski, Chris Gustin, Stephen Hughes, Christian Schneider, and Sven H\"{o}fling.
\newblock ``All-optical tuning of indistinguishable single photons generated in three-level quantum systems''.
\newblock \href{https://dx.doi.org/10.1021/acs.nanolett.1c04700}{Nano Letters {\bf 22}, 3562--3568}~(2022).

\bibitem{PhysRevA.76.042319}
Jens Koch, Terri~M. Yu, Jay Gambetta, A.~A. Houck, D.~I. Schuster, J.~Majer, Alexandre Blais, M.~H. Devoret, S.~M. Girvin, and R.~J. Schoelkopf.
\newblock ``Charge-insensitive qubit design derived from the cooper pair box''.
\newblock \href{https://dx.doi.org/10.1103/PhysRevA.76.042319}{Phys. Rev. A {\bf 76}, 042319}~(2007).

\bibitem{Roth2023}
Thomas~E. Roth, Ruichao Ma, and Weng~C. Chew.
\newblock ``The transmon qubit for electromagnetics engineers: An introduction''.
\newblock \href{https://dx.doi.org/10.1109/map.2022.3176593}{IEEE Antennas and Propagation Magazine {\bf 65}, 8--20}~(2023).

\bibitem{SnchezBurillo2015}
Eduardo S\'anchez-Burillo, Juanjo Garc\'ia-Ripoll, Luis Mart\'in-Moreno, and David Zueco.
\newblock ``Nonlinear quantum optics in the (ultra)strong light-matter coupling''.
\newblock \href{https://dx.doi.org/10.1039/c4fd00206g}{Faraday Discussions {\bf 178}, 335--356}~(2015).

\bibitem{Istrati2020}
D.~Istrati, Y.~Pilnyak, J.~C. Loredo, C.~Anton, N.~Somaschi, P.~Hilaire, H.~Ollivier, M.~Esmann, L.~Cohen, L.~Vidro, C.~Millet, A.~Lemaitre, I.~Sagnes, A.~Harouri, L.~Lanco, P.~Senellart, and H.~S. Eisenberg.
\newblock ``Sequential generation of linear cluster states from a single photon emitter''.
\newblock \href{https://dx.doi.org/10.1038/s41467--020--19341--4}{Nature Communications~{\bf 11}}~(2020).

\end{thebibliography}

\end{document}